\documentclass[aps,prl,reprint,groupedaddress]{revtex4-2}

\usepackage{color}
\usepackage{graphicx}
\usepackage{dcolumn}
\usepackage{bm}
\usepackage{amsmath}
\usepackage{ascmac}

\begin{document}

\title{Topological antiqued mechanical toy}

\author{Hirofumi Wada, Hayato Mizobata, Shuto Ueno and Taiju Yoneda}
\affiliation{Department of Physical Sciences, Ritsumeikan University, Kusatsu 525-8577, Japan}

\date{\today}

\begin{abstract}
{\it Jacob's ladder}---a classic children's toy---is a simple mechanical frame comprising rigid blocks connected by strings that shows curious unidirectional flipping waves. 
Nonetheless, its physical origin remains elusive. 
By combining experiment, numerical simulation, and theory, we show that understanding the underlying design principle of this toy requires diverse physical ideas. 
First, we conduct a water-tank experiment that excludes the collision-based mechanism, thus defying widespread expectations. 
Subsequently, we analytically demonstrate that the toy is bistable under gravity, thus implying its kink wave as a class of topological solitons.
The waves are surprisingly reminiscent---both experimentally and theoretically---to those in the Kane--Lubensky topological chain, owing to the stiffening of zero modes by the pretension under gravity.
However, a close examination based on the index theorem reveals that the similarity remains superficial and that the floppiness of the toy underlies the kink and antikink coexistence---a forbidden mode in the topological chain. 
By analyzing a generalized asymmetric toy, we reveal that its symmetric connection renders it topologically singular, thus resulting in amusing motions. 
We demonstrate these ideas by experimentally observing a dramatic pair annihilation of kink and antikink waves.
\end{abstract}

\maketitle

\begin{figure}[b]
\begin{center}
\includegraphics[width=0.99\linewidth]{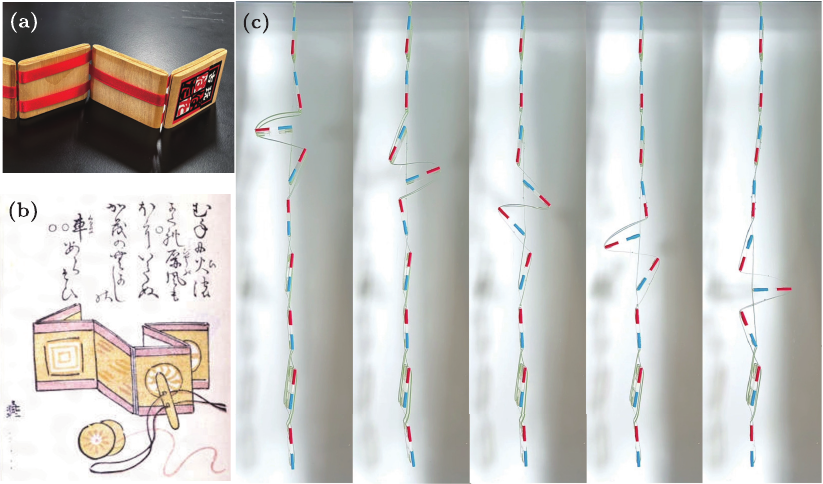}
\caption{Jacob's ladder model and experiment:
(a) Commercially available wooden toy; (b) photograph of the {\it hidden folding screen} from the Japanese toy encyclopedia {\it Edo Nishiki} published in the Edo period;
(c) water-tank experiment using a custom-designed model of $N=10$ acrylic blocks.
Kink separating the regions of the different orientations propagates down the ladder. Time proceeds from left to right. See also Supplemental Movie 2~\cite{SM}.}
\label{fig:experiment}
\end{center}
\end{figure}

{\it Introduction--}
Children's toys, such as spinning tops~\cite{Klein-Book}, rattlebacks~\cite{Bondi-PRLA-1986,Kondo-PRE-2017}, Euler's disk~\cite{Moffatt-Nature-2000}, and Newton's cradle~\cite{Sekimoto-EPJB-2021}, are important toolkits to visualize the concepts of physics, while simultaneously serving as a source of inspiration for new physical~\cite{Tokieda-Book-2026} and engineering problems~\cite{Bhamla-NatBiomed-2017}.
As an antique mechanical toy, {\it Jacob's ladder} features a periodic mechanical frame comprising wood blocks connected with tapes or ribbons (see Fig.~\ref{fig:experiment} (a))~\cite{Edge-PhysTeach-1998}.
In his short story included in {\it Household Words}, Charles Dickens described this toy as "made of little squares of red wood, that went flapping and clattering over one another, $\dots$ was a mighty marvel and a great delight"~\cite{Dickens, cotsen-blog}. 
In 1889, Scientific American described it as "The simple toy $\dots$ is very illusive in action"~\cite{SciAm-1889}.
Notably, the toy called "hidden folding screen" was described in a Japanese toy encyclopedia {\it Edo Nishiki} first published in 1773 (see Fig.~\ref{fig:experiment} (b))~\cite{Book-Edo}. 
When the ladder is suspended vertically and the top block is rotated by 180$^{\circ}$, the blocks rotate one by one until the bottom, with all pieces being flipped over. 
If the top block is re-rotated by 180$^{\circ}$ in the direction opposite to the first rotation, the wave is regenerated, and the toy returns to its original configuration~\cite{Edge-PhysTeach-1998}.
See Supplemental Movie 1~\cite{SM}.
While the configurational change of each block is trivially time reversible, the wave is unidirectional; it propagates consistently from top to bottom, thus breaking a time-reversal symmetry. 

In this study, we combine experiments, numerical simulations, and analytical theory to understand the physical principles underlying the mysterious movement of this puzzling toy.
We perform a water-tank experiment in reduced gravity, which suggests the illusive flipping cascade as a transition wave in a periodic system of bistable units.
By generalizing the actual toy design to an asymmetric one, we analytically and numerically show that the wave belongs to a class of topological kink solitons.
Subsequently, we argue its similarity (and dissimilarity) with the celebrated one-dimensional (1D) topological Maxwell chains, thereby highlighting the topological basis underlying the perplexing flipping behavior of the toy.

{\it Index Theorem--}
We begin with constraint counting in 2D.
We consider a ladder toy as a periodic arrangement of $N$ blocks with two stiff springs (or linkages). 
By construction, neighboring blocks rotate consistently in the opposite directions (see Supplemental Movie 1-3).
A free rigid block has three degrees of freedom (DoFs), and the total number of constraints is $N_b=2N-2$.
Based on the Calladine--Maxwell index theorem, $\nu=3N-N_b=N+2=N_0-N_s$, where $N_0$ is the number of zero modes (ZMs) and $N_s$ is the number of self-stress states (SSS)~\cite{Maxwell-PhilosMag-1864, Calladine-IJSS-1978}.
Thus, the number of mechanisms, or floppy modes, is $M=N_0-3=N-1+N_s$, indicating that the toy has at least $N-1$ mechanisms.
In fact, this significant floppiness underlies the mysterious kinematics of the toy.

{\it Experiment--}
To eliminate inertia, we conducted an experiment in a large tank $(27\, {\rm cm}\times 27\, {\rm cm} \times 80\, {\rm cm})$ that was filled with tap water (see SM Fig.~S6~\cite{SM}).
A toy was manually constructed, which comprised $N=10$ acrylic blocks (plus one supplemental top block) featuring a square face with side length $s=5$ cm and thickness $t=3$ mm. 
These blocks were connected to each other at a distance of $b=7.5$ cm using OPP film tapes measuring $40\, \mu$m thick.
Red and blue indicators were placed at each end of a block to define its orientation.
Because of buoyancy, the effective gravity becomes $g^{\ast}\approx 0.1 g$.
For further experimental details, see the SM~\cite{SM}.
By rotating the top block above the water surface, the flipping wave is created and propagates much more slowly than it would in air, 
with the speed $c_{\rm kink}=6.0$ cm/s under $b/s=1.5$ (see Fig.~\ref{fig:experiment} (c) and Movie 2)~\cite{SM}.
Notably, unlike the case in air, the flipping blocks never collide with the next block, seemingly different from a domino~\cite{Stronge-PRSLA-1987}.
While the clattering sound of the toy is pleasing, as described by Dickens ~\cite{footnote-1}, it is in fact not essential for the toy to function as designed.
\textcolor{black}{Nevertheless, a Jacob's ladder still shares its mechanism with the domino-like one as it develops a transition wave in a bistable energy landscape.}
In Fig.~\ref{fig:experiment} (c), the wave front defines the domain wall between the regions of the opposite orientation of the blocks.
We observed that the kink speed is insensitive to the initial excitation, unlike the case of energy-conserving systems, thereby suggesting that the frictional force balances the (reduced) gravitational driving.

\begin{figure}
\begin{center}
 \includegraphics[width=0.92\linewidth]{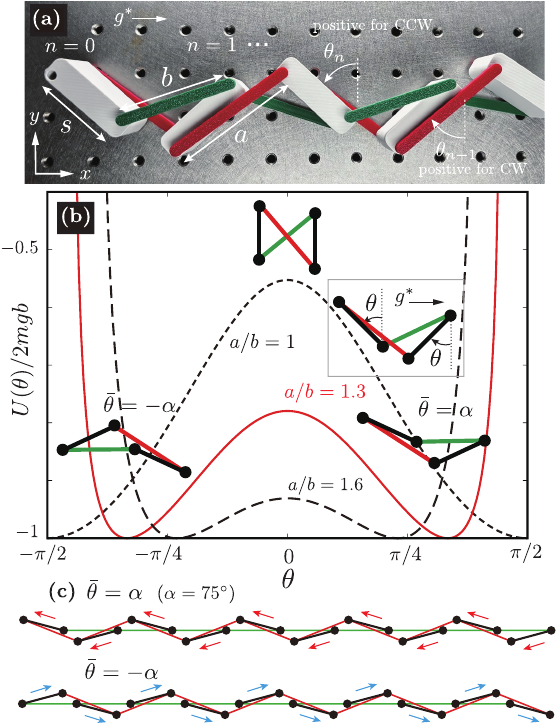}
\caption{Generalized asymmetric Jacob's ladder model:  
(a) 3D printed rigid model with red and green linkages of different lengths $a$ and $b$, where $a>b$.
\textcolor{black}{(b)} Net gravitational potential energy $U$ for the structure (shown in the insets) as a function of the zig-zag angle $\theta$ for various $a/b$, with $b/s=1.2$.
\textcolor{black}{(c)} Two equilibrium configurations \textcolor{black}{for $\bar{\theta}=\pm\alpha$, $\alpha=75^{\circ}$} and $N=10$. The arrow indicates the orientation of each block.
}
\label{fig:model}
\end{center}
\end{figure}

{\it Asymmetric ladder--}
Next, we consider an asymmetric model of $N$ unit cells, each of which comprises a rigid block and two springs of natural lengths $a$ and $b$, where $a\geq b$ \textcolor{black}{[Fig.~\ref{fig:model} (a)]}. 
For commercially available Jacob's ladders as well as the experiment above, the symmetry $a=b$ holds.
 
Although the toy may assume arbitrary configurations as it is floppy, the gravitational pulling uniquely determines the two zig-zag configurations. 
\textcolor{black}{To illustrate, let us consider an isolated criss-cross structure consisting of two blocks connected by two links [inset in Fig.~\ref{fig:model} (b)]. }
In the linkage limit, its effective gravitational potential \textcolor{black}{relative to the center of the top block} is expressed as
 \begin{equation}
 U(\theta) = -mg^{\ast} \sqrt{2(a^2+b^2)-4s^2\cos^2\theta-\frac{(a^2-b^2)^2}{4s^2\cos^2\theta}},
 \label{eq:U_G}
\end{equation}
where $\theta$ is \textcolor{black}{the angle common to the two blocks, as defined in the inset in Fig.~\ref{fig:model} (b)}.
\textcolor{black}{More precisely, for the periodic ladder studied below, the angle $\theta_n$ is defined as positive counterclockwise for even-$n$ blocks and positive clockwise for odd-$n$ blocks, 
where $n=0,1,\cdots,N-1$.
See also Fig.~\ref{fig:model_geometry}}.
Minimization of $U(\theta)$ yields \textcolor{black}{ two equilibrium angles, $\bar{\theta}=\alpha (>0) $ and $-\alpha$, where $a^2 = b^2 +4s^2\cos^2\alpha$}. 
\textcolor{black}{The two ground-state configurations [Fig.~\ref{fig:model} (c)] are} similar to those assumed in the Kane--Lubensky (KL) topological chain model with lattice spacing $b$ and spring length $a$~\cite{Kane-NatPhys-2014, Lubensky-Review-2015}.
Nevertheless, the long straight bar in the KL chain is replaced in this study with $N$ linkages of length $b$, thus providing additional $N$ DoFs. 
Figure~\ref{fig:model} (b) shows the generic bistability with the two energetically degenerated block orientations (up and down) for different values of $a/b$.
In the symmetric limit $a/b\rightarrow 1$, we obtain $\alpha\rightarrow \pi/2$, which corresponds to the vertically aligned configuration of an actual toy.
\textcolor{black}{For convenience, we will also use $r=s/2$ in the following.}

Next, to visualize the vibrational spectrum around the zig-zag configuration stabilized by the gravity, we replace the linkages with stiff springs of stiffness $k$ under the periodic boundary conditions (PBCs).
In 2D, the center of mass of the $n$-th block has two translational and one rotational DoF, as described by the displacement vector $\delta{\bm x}_n=(\delta u_n,\textcolor{black}{(-1)^n}\delta v_n,\delta\theta_n)$ for \textcolor{black}{$n=0,1,\cdots,N-1$}.
Under the PBCs, we shift to the Bloch space by Fourier transforming all the relevant quantities
and work with the wavenumbers $q$ in the first Brillouin zone $-\pi/b \leq q\leq \pi/b$.
In the linear analysis, the spring extensions $(e_{A,n},e_{B,n})$ are related to $\delta{\bm x}_n$ as $(e_A(q),e_B(q))={\bf C}(q)\cdot \delta {\bm x}(q)$~\cite{Lubensky-Review-2015}, where the compatibility matrix ${\bf C}(q)$ is the $2 \times 3$ matrix given by
\begin{equation}
 {\bf C}(q) =
\left(
\begin{array}{ccc}
  (b/a)(e^{iqb}-1) &  \textcolor{black}{4\gamma (e^{iqb}+1)}  & -c_1e^{iqb}+c_2 \\
  e^{iqb}-1 &  0 &  \gamma a (e^{iqb}-1), 
\end{array}
\right),
\label{eq:C-q}
\end{equation}
where we define $\gamma = (s/2a)\cos\alpha$, $c_1 = \gamma (b+2s\sin\bar{\theta})$, and $c_2 = \gamma (b-2s\sin\bar{\theta})$.
The ZMs are the null space vectors of ${\bf C}(q)$.
An explicit analytical result that ensures $n_0(q)=1$ \textcolor{black}{for $q\neq 0$ and $n_0(q=0)=2$}, where $n_0(q)$ denotes the number of mode-$q$ ZMs, is provided in the SM~\cite{SM}.
\textcolor{black}{We therefore have $N_0=\sum_{q} n_0(q) =N+1$}.
The equilibrium matrix, ${\bf Q}(q)$, expressed based on force and moment balance conditions, is derived from ${\bf Q}(q)={\bf C}^{\dagger}(q)$.
No SSS exists for $q\neq 0$ and $a/b>1$, i.e., $n_s(q)=0$, \textcolor{black}{but one SSS exists for $q=0$ (because of the straight green linkages in Fig.~\ref{fig:model} (c)); thus, $N_s=1$. }
Therefore, we confirm that $N_0-N_s=N$, as required by the Index theorem.
\textcolor{black}{Additionally, all these ZMs are likely to be finite mechanisms~\cite{SM}.}

Meanwhile, the kinematics of {\it symmetric} Jacob's ladder differs significantly.
For $a=b$, we have $\gamma=0$, and $c_1=c_2=0$ in Eq.~(\ref{eq:C-q}).
The ZMs and SSS eigenvectors are expressed as
\begin{equation}
{\rm ZM}  = \epsilon_1(q)\left(
\begin{array}{c}
  0 \\
  1 \\
  0   
\end{array}
\right)
+
\epsilon_2(q) \left(
\begin{array}{c}
  0 \\
  0 \\
  1   
\end{array}
\right),
\ 
{\rm SSS} =
t(q)\left(
\begin{array}{c}
  1 \\
  -1   
\end{array}
\right), 
\end{equation}
where $\epsilon_{1,2}(q)$ and $t(q)$ are arbitrary (yet sufficiently small) parameters. 
This results in $n_0(q)=2$ and $n_s(q)=1$, which satisfy the Index theorem $n_{\rm free}-n_B=1=n_0(q)-n_s(q)$.
The symmetric toy thus contains more inifinitesimal mechanisms, i.e., $M=2N-2$.

To establish a connection to the topological chains, we now consider constraining some of the floppy modes. 
Specifically, we impose \textcolor{black}{$\delta v_n = -r \sin\bar{\theta}\, \delta\theta_n$}. 
Under PBC, these $N$ additional constraints increase $N_b$ by $N$, and the index count yields $\nu=3N-2N-N = 0$.
Thus, the toy becomes a Maxwell frame, as confirmed by the fact that ${\bf C}(q)$ becomes a square ($2\times2$) matrix, expressed as
\begin{eqnarray}
 {\bf C}(q) &=&
\left(
\begin{array}{ccc}
  (b/a)(e^{iqb}-1) & -v_1e^{iqb}+v_2 \\
  e^{iqb}-1 & \gamma a (e^{iqb}-1) 
\end{array}
\right),
\label{eq:C_red}
\end{eqnarray}
where we define $v_1 = c_1+2\gamma s \sin\bar{\theta}$ and $v_2 = c_2 -2\gamma s \sin\bar{\theta}$.
\textcolor{black}{By examining ${\bf Q}(q)$, we again observe one SSS for $q=0$, i.e., $N_s=1$ and thus $N_0=1$ from the index theorem, and no mechanism $M=N_0-1=0$.}

In fact, the additional constraints specified above are clearly supported experimentally. 
When the actual Jacob's ladder is subjected to vertical pulling by gravity, the tension stabilizes most of the infinitesimal mechanisms (or bulk ZMs); the dispersion of phonons for transverse deflections is gapped effectively for $|q|>0$.
The constraints on $\delta v_n$ above are an idealization of this geometric stiffness, in a sense that all the ZMs disappear due to the pre-stress under gravity~\cite{Pellegrino-IJSS-22, Pellegrino-IJSS-1990, Guest-IJSS-2005}. 
In other words, a toy with $a>b$ is a {\it softly} 1-DoF system under gravity, which possesses the same mechanical connectivity and stability with the KL chain but may have more DoFs for nonlinear deformations.
 
For a free ladder toy cut out from a periodic one, two springs are removed; therefore, we have $\nu=2$ from the index count.
\textcolor{black}{Because the $q=0$ SSS in PBC has been lost}, we have $N_s=0$, leading to $N_0=2$ when we assume no localized SSS (i.e., no domain wall).
Solving $e_{A,n}=e_{B,n}=0$ in terms of $\delta{\bm x}_n$ yields two ZMs.
The first one is the trivial rigid body translation (along $x$), whereas the second one represents surface ZM that decays in the bulk, expressed as
\begin{equation}
 \delta\theta_{n+1} = \frac{c_2}{c_1} \delta \theta_n, 
 \
 \delta u_{n+1}-\delta u_n = \textcolor{black}{\frac{s^2\sin(2\bar{\theta})}{b+2s\sin\bar{\theta}}} \delta \theta_n.
 \label{eq:2nd_sol}
\end{equation}
This ZM localizes exponentially at the edge as $\delta\theta_n = \delta\theta_0 e^{-\kappa n b}$,
where the inverse of the penetration length is expressed as
\textcolor{black}{$\kappa = b^{-1}\log |(b+2s\sin\bar{\theta})/(b-2s\sin\bar{\theta})|$}, which is consistent with the ZM identified in the KL model~\cite{Lubensky-Review-2015}.
The conclusion above is valid for $|c_2/c_1|<1$, i.e., \textcolor{black}{$0<\bar{\theta}\leq \pi/2$}. 
For \textcolor{black}{$\bar{\theta}<0$}, we obtain $\delta\theta_{n+1}/\delta\theta_n >1$ from Eq.~(\ref{eq:2nd_sol}), which implies that 
the surface ZM should localize at the bottom ($n=\textcolor{black}{N-1}$) edge rather than at the top.
The length $\kappa^{-1}$ diverges at $\bar{\theta}\rightarrow 0$, i.e., at the unstable fixed point [Fig.~\ref{fig:model} (b)].
The same conclusion can be inferred by considering a topological winding number associated with the determinant of ${\bf C}(q)$.
Additionally, the topological polarization defined according to Ref.~\cite{Lubensky-Review-2015} simply coincides with the $x$-projection of the orientation of each block.
However, again, for the symmetric design $a=b$, the circle on the complex plane $\det|{\bf C}(q)|$ contracts towards the origin, thus rendering its topological phase undefined within the present linear analysis.
See the detailed argument presented in the SM~\cite{SM}.

\begin{figure}[htbp]
\begin{center}
 \includegraphics[width=0.99\linewidth]{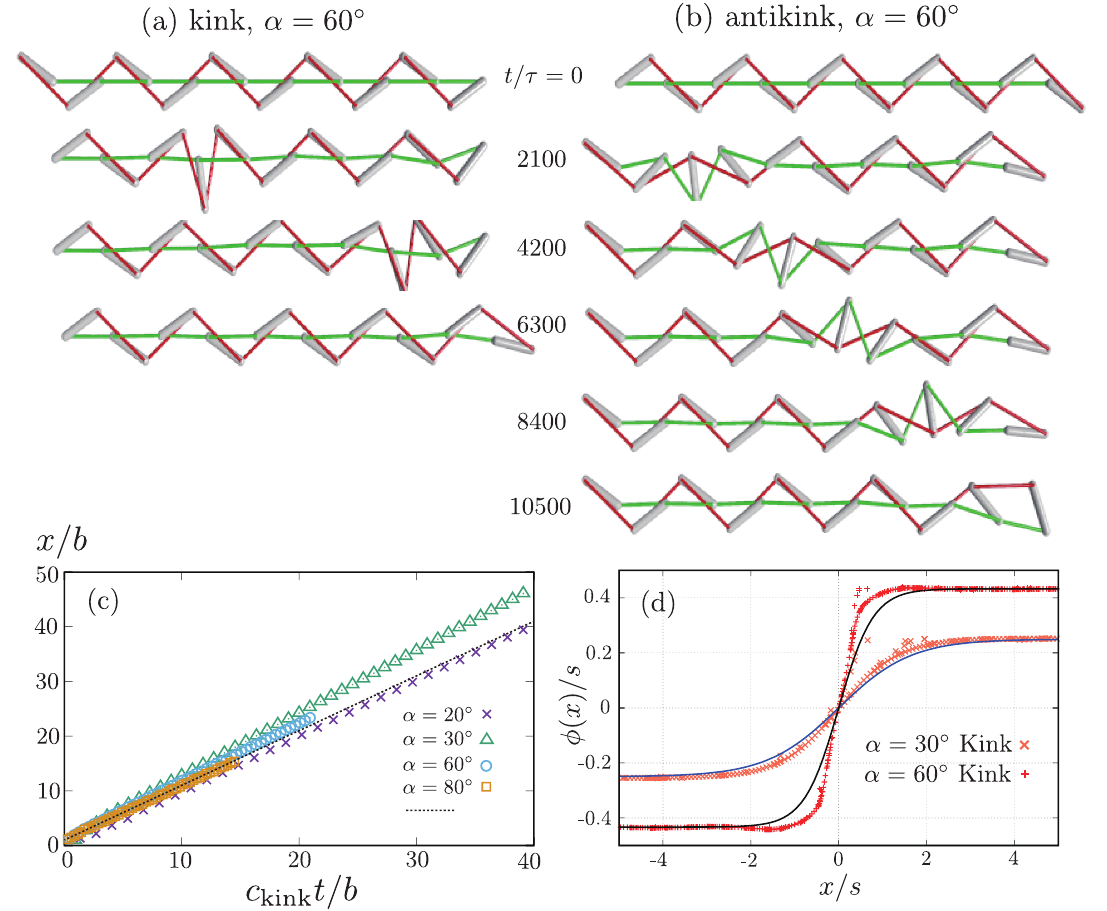}
\caption{Finite amplitude waves in numerical simulations: (a) kink and (b) antikink for $N=10$, $\alpha=60^{\circ}$ and $b/s=1.3$.
\textcolor{black}{(c) Kink position as a function of rescaled time $c_{\rm kink}t/b$, with $c_{\rm kink}$ predicted in Eq.~(\ref{eq:c_kink}) for various $\alpha$ and $N=50$.
(d) Shapes of a propagating kink for $\alpha=30^{\circ}$ and $60^{\circ}$ ($N=50$). The shapes at different times are shifted so they collapse.
The solid lines are the theoretical prediction of $\tanh$ given in the main text with the parameters used in the corresponding simulations.} 
The unit of time $t$ is $\tau=\sqrt{m/k}$, where \textcolor{black}{$2m$} is the block mass.}
\label{fig:solitons}
\end{center}
\end{figure}

Nevertheless, the toy is essentially floppy.
This allows nonlinear large-amplitude waves \textcolor{black}{that may be richer than those} in an exact 1-DoF topological chain.
To visualize the number of evolving finite mechanisms in the ladder, we investigate the full nonlinear dynamics by using the numerical method described in the SM~\cite{SM}.
Numerically, we discovered that the toy supports a large-amplitude antikink, in addition to a kink (see Figs.~\ref{fig:solitons}) \textcolor{black}{for large $\alpha$}.
An initial configuration can be either one of the two ground states, and the first block is rotated by $\pm 2\alpha$, where its angle changes as $\pm \alpha \rightarrow \mp \alpha$.
For \textcolor{black}{$\bar{\theta}=60^{\circ}$}, a ZM at the domain separating the regions with opposite orientations evolves into a propagating kink [see Fig.~\ref{fig:solitons} (a) and \textcolor{black}{SM Movie 4}]\cite{SM}.
During the propagation, the shorter (green) springs remain almost horizontal; the kink resembles the flipper soliton in the KL chain~\cite{Chen-PNAS-2014}.

In contrast to the KL chain, the antikink can propagate into the bulk \textcolor{black}{for a large $\alpha$} coupled with the zigzag deformations of the shorter (green) springs [see Fig.~\ref{fig:solitons} \textcolor{black}{(b)} and \textcolor{black}{SM Movie 5}]~\cite{SM}.
When the antikink reaches the bottom, the toy recovers its original configuration, except for the final block.
(Compare (a) $t/\tau=0$ and \textcolor{black}{(b) $10500$.})
For the symmetric design $a/b \rightarrow 1$, i.e., \textcolor{black}{$\alpha\rightarrow 90^{\circ}$}, the \textcolor{black}{edge} effect is eliminated and the forward and backward flippings become precisely cyclic, as demonstrated in the simulation for \textcolor{black}{$\alpha=87^{\circ}$ (SM Movie 6 and 7)~\cite{SM}.
Only in this limit, the domain walls become ZMs for the both cases.}
This is how the Jacob's ladder creates a visual illusion of the blocks cascading down the strings endlessly.
\textcolor{black}{For $\alpha\lesssim 50^{\circ}$, antikink behavior is found to be more involved due to SSS (see SM Fig.~S7 and Movies 9 and 10~\cite{SM}). Further studies of this topic are ongoing.} 

\textcolor{black}{
We developed a continuum theory for the kinks (see SM~\cite{SM}). The scalar field $\phi(x,t)=r\sin\theta(x,t)$ obeys the dynamic equation given by
\begin{equation}
 \frac{I}{r^2}\ddot{\phi} = -\frac{4\zeta}{b}\dot{\phi}+\tilde{\mu}\phi^{\prime\prime}-\mu V(\phi)\frac{dV(\phi)}{d\phi}-\frac{2mg^{\ast}}{b},
 \label{eq:eqm-kink}
\end{equation}
where $I=ms^2/b$ is the moment of inertia per length, $\dot{\phi}=\partial_t\phi$, $\phi^{\prime}=\partial_x\phi$, $\mu=kb$, $\tilde{\mu}=\mu(1+b^2/a^2)$, $V(\phi)=(8/ab)(\phi_0^2-\phi^2)$, $\phi_0=r\sin\alpha$, and 
we assume $\phi^2/r^2\ll 1$~\cite{Chen-PNAS-2014}.
In Eq.~(\ref{eq:eqm-kink}), we assume that the frictional force is proportional to the local velocity of the nodes with coefficient $\zeta$.
In the viscous regime, where friction and gravity are balanced, we neglect the inertia term in Eq.~(\ref{eq:eqm-kink}) and 
obtain a steady-state solution for the weak driving regime $\epsilon= 4mg^{\ast}/(ks \sin\alpha)\ll 1$ as follows:
$\phi(x,t)/\phi_0=-(a/8s\sin\alpha)^2\epsilon+\tanh \{4s\sin\alpha (x-c_{\rm kink}t)/b\sqrt{a^2+b^2}\}$ [Fig.~\ref{fig:solitons} (d)], where the kink speed is predicted as
\begin{eqnarray}
 c_{\rm kink}(\alpha)  &=& \frac{3b\sqrt{a^2+b^2}}{8 s^2 \sin^2\alpha}\frac{mg^{\ast}}{\zeta}.
 \label{eq:c_kink}
\end{eqnarray}
While $c_{\rm kink}\sim mg^{\ast}/\zeta$ may be physically easy to understand, the predicted $\sim 1/\sin^2\alpha$ dependence appears non-trivial, as confirmed by the 
numerical data [Fig.~\ref{fig:solitons} (c)].
A continuum description of antikink is currently challenging and will be left to future studies. 
}

\begin{figure}
\begin{center}
 \includegraphics[width=0.99\linewidth]{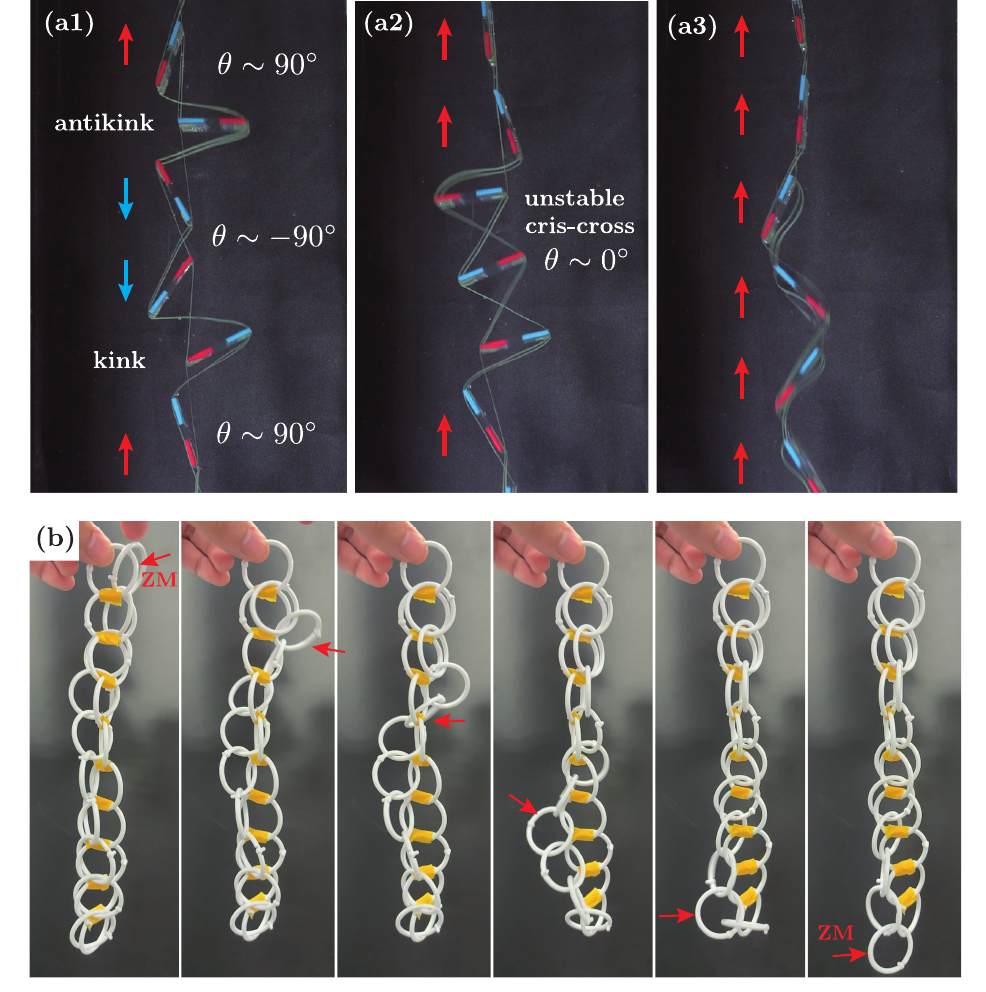}
\caption{Kink behaviors in two experimental toy systems:
(a) Pair annihilation of kink and antikink in water-tank experiment. Arrows indicate block orientations.
(a-1) kink and antikink before collision; (a-2) criss-cross at collision site; (a-3) long-wavelength deformations after annihilation.
(b) {\it Spin wave} propagating down a dual ring chain constraint. ZM initially localized at the top edge migrates helically toward the bottom.
See also Supplemental Movies 12--14~\cite{SM}.}
\label{fig:exp-toys}
\end{center}
\end{figure}

The above numerical results suggest that kink and antikink coexist for large $\alpha$, which we now confirm experimentally.
As shown in Fig.~\ref{fig:exp-toys} (a-1), we successfully generated a kink-and-antikink pair (Movie 11)~\cite{SM}.
Remarkably, the antikink can approach near the kink, thus resulting in a dramatic pair annihilation (Movie 12)~\cite{SM}. 
Upon the collision, an unstable criss-cross configuration, where the symmetry is recovered, is observed [see Fig.~\ref{fig:exp-toys} (a-2)], which is consistent with the bistability shown in Fig.~\ref{fig:model} (b).
\textcolor{black}{The unexpected asymmetry in the wave speeds, $c_{\rm antikink}/c_{\rm kink}>1$, arises from the limitation of our fabrication precision; preferential curvatures in the tapes at the hinges may slightly deform the bistable energy landscape, with one stable configuration having a lower energy than that of the other. See SM for details~\cite{SM}.
We can also demonstrate a head-on collision between the kink and antikink by hanging a ladder and creating waves at both ends in the air (Appendix B).}

In summary, we investigated the kinematics and mechanics underlying the perplexing movements of an antique toy, i.e., Jacob's ladder.
Based on the Calladine--Maxwell Index theorem, we demonstrated that the toy is designed essentially floppy.
However, because the toy functioned as intended in gravity, it became stiff owing to gravity-induced pre-tension.
\textcolor{black}{While the floppiness intrinsically allows any number of waves along either direction, gravity simply selects them.
This mechanism is essentially extrinsic and not fully protected by topological polarization.}
The exquisite design allows Jacob's ladder to support the stable unidirectional propagation of both the kink and antikink, which is the key feature of the seemingly endless, illusive flippings described by Dickens.
 
Finally, the proposed scenario became more evident for a double-stranded ring chain shown in Fig.~\ref{fig:exp-toys} (b). \textcolor{black}{See also Appendix C.}
By releasing the top ring, we observed a helical solitary wave, in which the rings in one strand flipped sequentially~\cite{Edge-PhysTeach-1998}.
The wave was observed most clearly when the partner strand was constrained to reconfigure with the yellow tapes shown in Fig.~\ref{fig:exp-toys} (b). 
\textcolor{black}{See also SM Movie 13~\cite{SM}.}
In this case, when the ZM at the top reaches the bottom, one must ensure that the whole chain is upside down to repeat the wave propagation. 
Undoubtedly the actual toy functions differently. 
The two-strand chains were equally floppy; by retrieving a top ring of one of two strands alternately, one can generate a unidirectional wave indefinitely, exactly as in Jacob's ladder~\cite{comment-unpublished}. 

In recent years, new types of solitary waves have been extensively investigated for various periodic mechanical systems~\cite{Deng-JAP-2021}, including origami~\cite{Yasuda-SciAdv-2019, Imada-PRRes-2025}, lattices composed of rotating~\cite{Deng-PRL-2017} or snapping units~\cite{Nadkarni-PRL-2016}, metastable woven frames~\cite{Boucher-PRL-2017} and nonreciprocal metamaterials~\cite{Veenstra-Nature-2024}, thus significantly expanding our understanding of solitary waves in classical nonlinear lattices~\cite{Scott-AmJPhys-1969, Toda-book, Berman-Chaos-2005} and continuum media~\cite{Kawasaki-PhysicaA-1982}.
The antique toy investigated in this study has been around since ancient times; nonetheless, its nonlinear waves possess physical and topological properties that may be worth investigating in emerging fields.
The toy-inspired wave-control design in this study offers a valuable avenue for harnessing and controlling large-amplitude waves for functionalities in architected materials, soft robotics, and mechanical diodes, rectifiers, and switches.
Additionally, the present study may serve as a unique conceptual guidance, at least metaphorically, to a range of dynamic biophysical problems such as the propagation of allosteric structural changes in proteins~\cite{Sekimoto-EPL-2024} and the enigmatic mechanism of swimming bacteria using a kink-propagation pair based on the handedness of their helical bodies~\cite{Goldstein-PRL-2000, Shaevitz-Cell-2005, Nakane-JBacteriol-2020, Sasajima-FrontBiol-2021}.

\begin{acknowledgments}
We thank Dr. K. Yoshida for highlighting the traditional Japanese toy {\it Pata pata} (Jacob's ladder). 
We also thank Prof. L. H. Kauffman for introducing a chain of key rings that generates a helical wave 
during the "Periodic Tangle" Workshop in Sendai, Japan, in 2025.
We thank Prof. P. M. Reis and Dr. D. Shimamoto for their discussions and encouragement.
Financial support from JSPS KAKENHI (grant no. 23K22463 to H.W.) is acknowledged.
\end{acknowledgments}

%


\section{Appendices}
\paragraph*{Appendix A: Theoretical setup:}
\begin{figure*}[htb]
\begin{center}
 \includegraphics[width=0.78\linewidth]{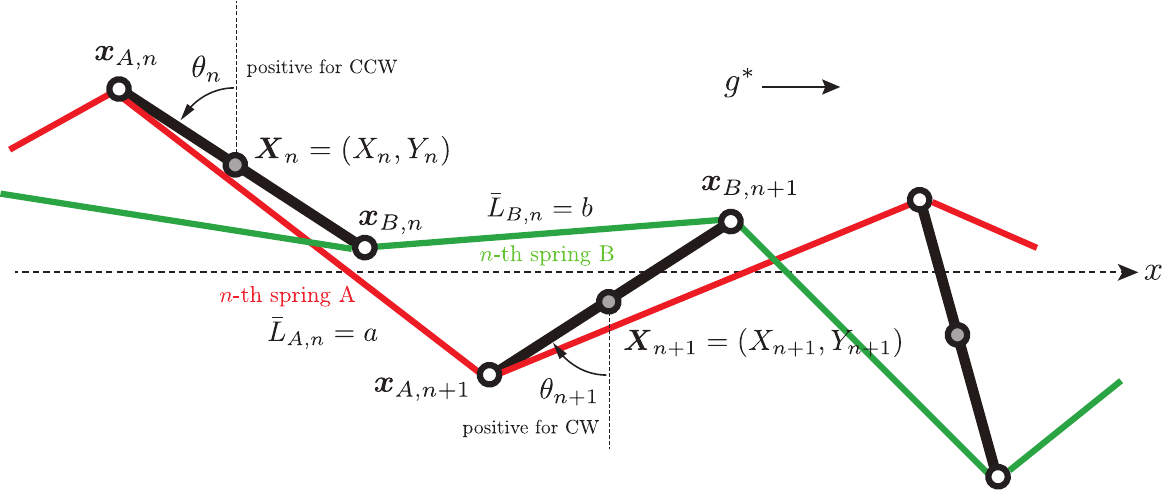}
\caption{\textcolor{black}{Definitions of the coordinate system, relevant variables, and parameters of a generalized asymmetric Jacob's ladder model. The rotational angle $\theta_n$ 
of the $n$-th rigid block is defined as positive CCW for even-$n$ blocks and positive CW for odd-$n$ blocks, where $n=0,1,\cdots,N-1$.}}
\label{fig:model_geometry}
\end{center}
\end{figure*}
\textcolor{black}{
Figure~\ref{fig:model_geometry} shows the coordinate system and definitions of the relevant variables and parameters used in our theoretical analysis.
The center-of-mass position of the $n$th block of length $s=2r$ is represented by ${\bm X}_n=(X_n,Y_n)$, where $n=0,1,\cdots, N-1$.
Additionally, a node connected to spring A (colored red) is denoted as ${\bm x}_{A,n}$, whereas a node connected to spring B (colored green) is ${\bm x}_{B,n}$.
We define a unit vector pointing from ${\bm x}_{A,n}$ to ${\bm x}_{B,n}$ as 
\begin{eqnarray}
 \hat{\bm t}_n = (\sin\theta_n, -(-1)^n\cos\theta_n),
\end{eqnarray}
where $(-1)^n$ denotes the underlying zig-zag configuration of the asymmetric ladder.
Using these defined quantities, we obtain ${\bm x}_{A,n}={\bm X}_n-r\hat{\bm t}_n$ and ${\bm x}_{B,n}={\bm X}_n+r\hat{\bm t}_n$.
}

\textcolor{black}{
By writing the spring length $L_{\sigma,n} = |{\bm x}_{\sigma, n+1}-{\bm x}_{\sigma, n}|$, where $\sigma=A, B$, the extensions can be written as 
$e_{\sigma, n} \simeq (L^2_{\sigma, n}-\bar{L}^2_{\sigma})/2\bar{L}_{\sigma}$, where $\bar{L}_A=a$ and $\bar{L}_B=b$ are natural lengths.
The total energy of the proposed system, $E$, is given by
\begin{eqnarray}
 E &=& \sum_{n=0}^{N-1} \frac{k}{2} (e_{A,n}^2+e_{B,n}^2) - 2mg^{\ast} X_n,
 \label{eq:E_tot}
\end{eqnarray}
where $m$ is the mass of nodes A and B. Thus, the total mass of one block is $2m$.
}

\begin{figure}[htb]
\begin{center}
 \includegraphics[width=0.97\linewidth]{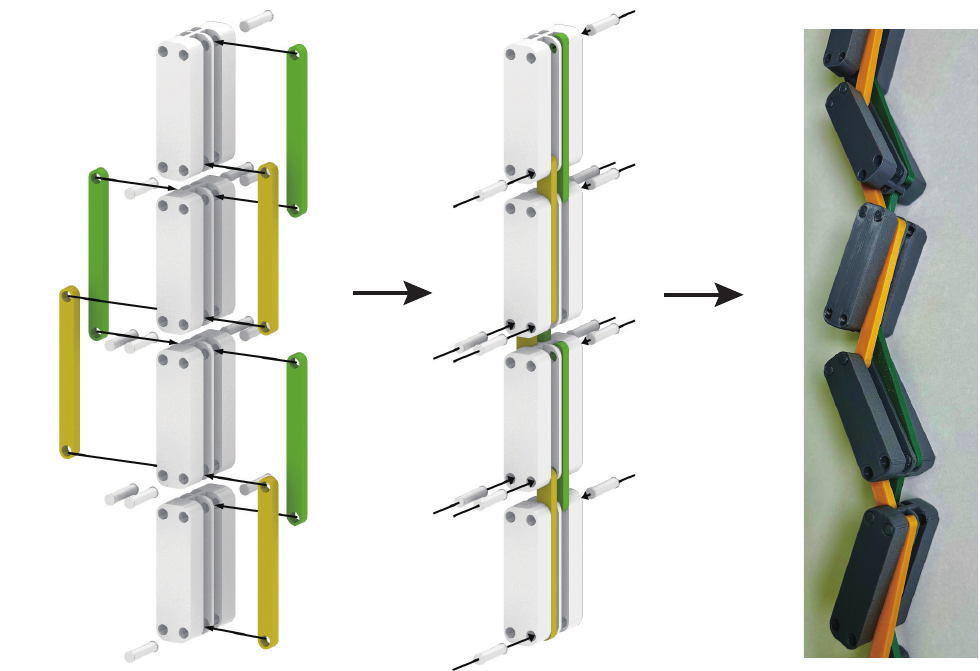}
\caption{\textcolor{black}{Schematic of the assembly of 3D-printed Jacob's ladder model with $a=b$.}}
\label{fig:print-model}
\end{center}
\end{figure}

\begin{figure*}[htb]
\begin{center}
 \includegraphics[width=0.97\linewidth]{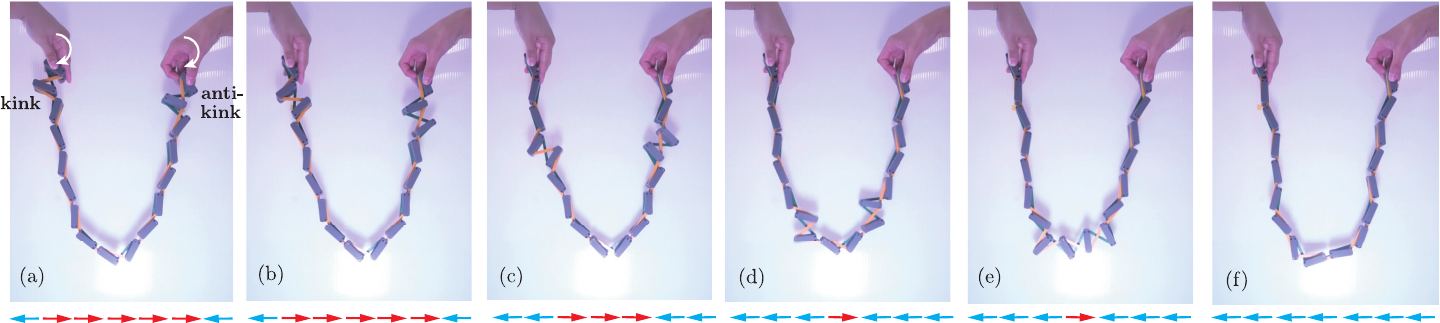}
\caption{\textcolor{black}{Demonstration of a head-on collision between kink and antikink in air. Time proceeds from (a) to (f), with an interval of approximately 0.2 seconds. 
A model shown in Fig.~\ref{fig:print-model} consists of $N=17$ blocks here. The row of red and blue arrows at the bottom of each subfigure schematically shows the flipping sequence of the orientations of the blocks, with the blue-red and red-blue boundaries defined as kink and antikink, respectively. See Supplemental Movie 14.}}
\label{fig:head-on}
\end{center}
\end{figure*}

\textcolor{black}{
\paragraph*{Appendix B: Head-on collision}
To demonstrate a head-on collision between a kink and antikink, we designed and fabricated a 3D-printed rigid-linked Jacob's ladder of $a=b$ as shown in Fig.~\ref{fig:print-model}.
This model operates much more smoothly than the model shown in Fig. ~\ref{fig:model} (a), allowing us to observe collisions in air.
When waves are created from both edges, they propagate toward the center and collide in a chain suspended in air. See Fig.~\ref{fig:head-on} and SM Movie 13~\cite{SM}.
Upon the collision shown in Fig.~\ref{fig:exp-toys} (a), the ladder returns to its original configuration.
Conversely, for the head-on collision addressed here, the ladder changes to the other ground state configuration, in which the orientations of all blocks are flipped.
This is schematically illustrated at the bottom of Fig.~\ref{fig:head-on}.
}

\paragraph*{Appendix C: Architecture of a double-ring chain}
\begin{figure}[htb]
\begin{center}
 \includegraphics[width=0.90\linewidth]{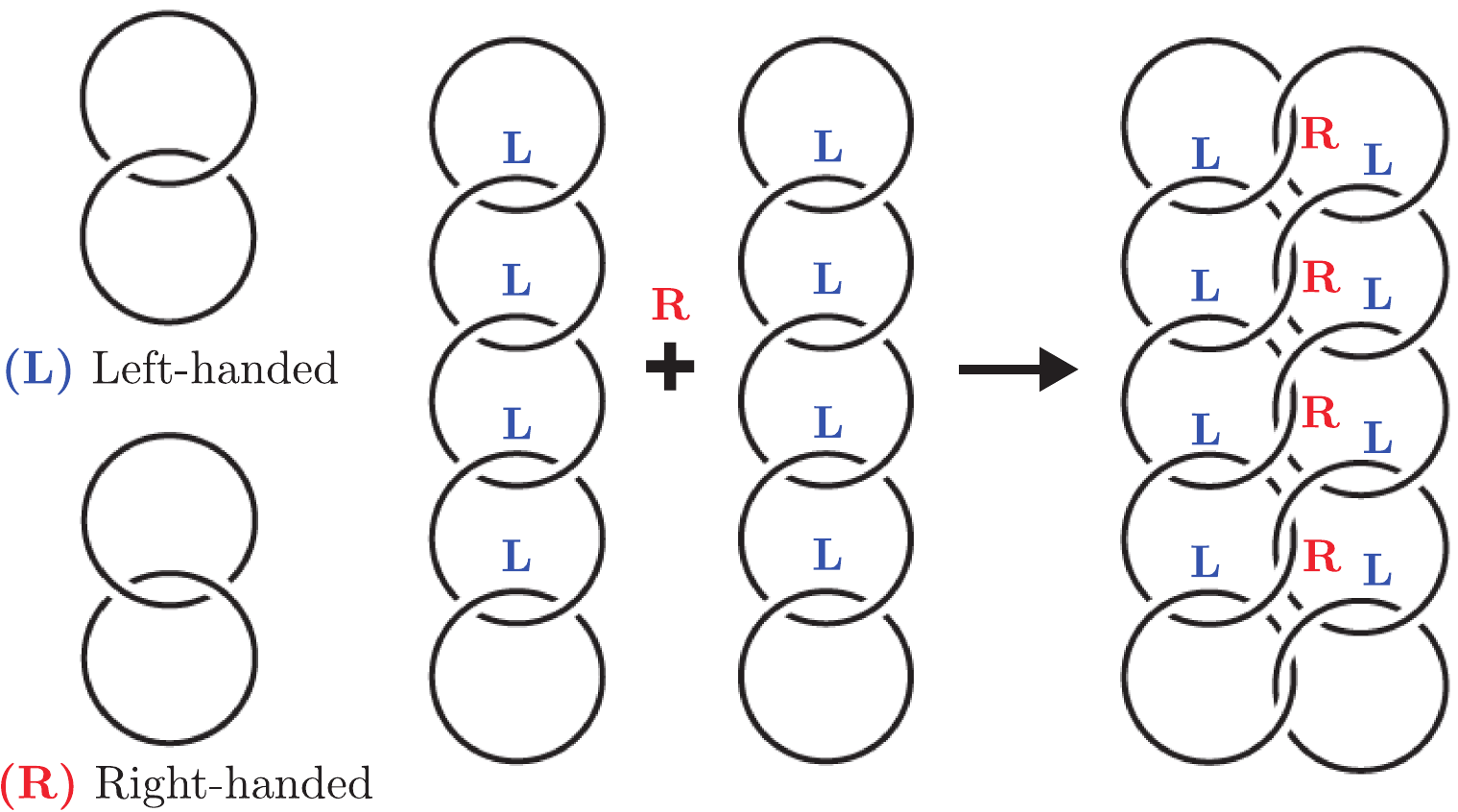}
\caption{\textcolor{black}{How to connect rings to form a double-ring chain shown in Fig.~4 (b). The connection shown here forms a left-handed doublet chain.
If all of R and L are simultaneously swapped, a right-handed doublet chain is formed.}}
\label{fig:ring}
\end{center}
\end{figure}
\textcolor{black}{
Figure~\ref{fig:ring} illustrates how the rings are connected to form a double-ring chain, as shown in Fig.~\ref{fig:exp-toys} (b). 
The two rings may be linked in one of two ways, which are defined as the right-handed (R) unit and the left-handed unit (L). 
Although these units can transform into each other when isolated, the units within the periodic structure cannot switch between R and L. 
Therefore, they are topologically distinct units.
Only the rings at the edges can rotate freely and carry ZMs.
These connections are illustrated in Fig.~\ref{fig:ring} and form a left-handed doublet chain along which a right-handed (right-circularly polarized) wave can propagate. 
Notably, the chain has chiral symmetry. 
If all R and L units are swapped simultaneously, a right-handed chain is formed and a left-handed (left-circularly polarized) wave propagates. 
Further details are provided in Ref.~\cite{unpublished}.
}

\onecolumngrid
\begin{center}
{\large \bf SUPPLEMENTAL MATERIALS}
\end{center}

\section{Jacob's ladder toy}
The toy known as Jacob's Ladder has numerous alternative names—Aaron's Bell, Chinese Block, Click-Clack Toy, Magic Tablet, and Tumbling Block—and much about its origins and history remains unclear. 
According to Wikipedia, the origins of Jacob's ladder could date back to ancient Egypt and/or ancient China. 
However, the credibility of such information appears to be rather low, according to the survey reported in Ref.~\cite{cotsen-blog}.
The same source provides a more convincing account of the toy's history—particularly its history in the United States, as well as how its value has been redefined today and the new roles that it now plays in various contexts including the Child developmental research.

A Jacob's ladder toy has also long been known in Japan as one of the representative folk mechanical toys. 
In particular, during the Edo period (1603-1868), it was known as a "changing folding screen" and appears to have been popular among people. 
It is said that the toy made in the Edo period consisted of six cedar boards connected with white Japanese paper~\cite{Karakuri-Book}.

\begin{figure}[htbp]
\begin{center}
 \includegraphics[width=0.60\linewidth]{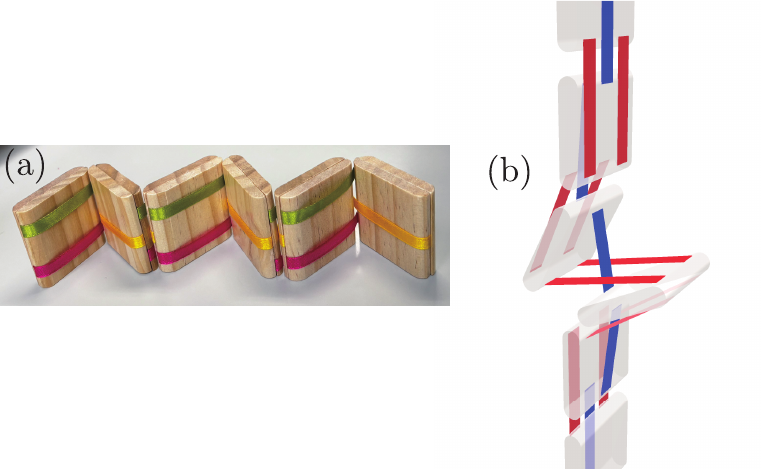}
\caption{(a) Photograph of commertial Jacob's ladder toy (Toysmith.com) (b) Computer graphic image of Jacob's ladder whose rigid blocks are partially transparent so that the arrangement of the ribbons on the backside is also visible.}
\label{fig:toy}
\end{center}
\end{figure}

Currently, Jacob's ladder toys available on the market commonly have the structure shown in the Fig.~\ref{fig:toy} (a). 
One may notice that the arrangement consists of a repeating pattern of two strings (or ribbons) followed by one. 
However, this arrangement is dictated by the constraints that adjacent strings must not overlap during the movement of the blocks and must be arranged symmetrically; the repeating pattern of two followed by one does not have any inherent significance in itself.

In fact, by replacing the ribbons with link mechanisms (rigid bars), one can find a different arrangement that satisfies the self-avoiding constraint as well as the symmetry. 
This is shown in Fig.~2 (a) in the main manuscript. 
Here, there is exactly one link connecting the same side of each block. 
This linkage provides the most suitable model for mathematical analysis of the mechanism of the Jacob's ladder, as we present in the main manuscript.
For $\alpha\simeq 90^{\circ}$ model, the ground state configuration is nearly stragiht, and we observe a smooth flipping wave under the gravity alone in air for both forward and backward rotations. 
For $\alpha\simeq 69^{\circ}$ model, the ground state configuration is a zig-zag one. 
\textcolor{black}{We observe the smooth kink wave for the forward rotation, but for the backward rotation, antikink wave seems more suble and complicated. 
Overall, these observation is in agreement with antikink behavior observed in our numerical simulation. Antikink moves slowly for $\alpha =60^{\circ}$, but it is oberved to stop for $\alpha\lesssim 50^{\circ}$, as will be described in Sec.~\ref{sec:simulation}.}
Even so, by using your hands to assist this rigid link toy's movement slightly, you can now figure out much more easily the mechanism by which Jacob's ladder supports the illusive flipping kink and antikink waves.

\section{Equilibrium configuration under gravity}
\begin{figure}[htbp]
\begin{center}
 \includegraphics[width=0.60\linewidth]{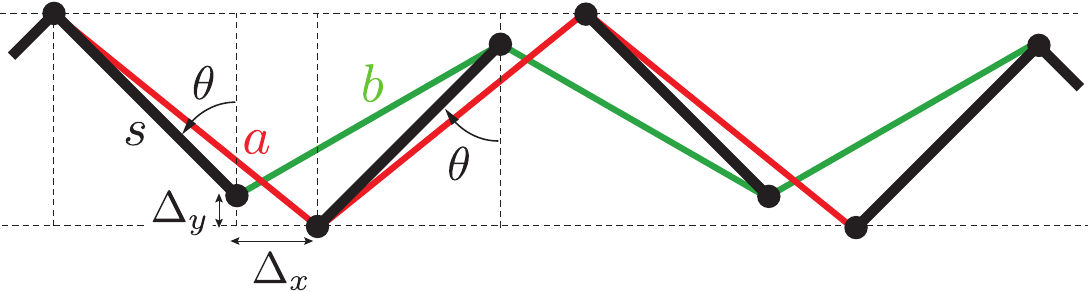}
\caption{Schematic of the model configuration for the calculation of $\Delta_x(\theta)$ and $\Delta_y(\theta)$.}
\label{fig:sm_model}
\end{center}
\end{figure}

\begin{figure}[htbp]
\begin{center}
 \includegraphics[width=0.50\linewidth]{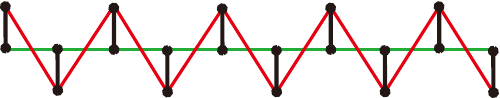}
\caption{Gapless high symmetry equilibirum configuration for $\alpha=0^{\circ}$ and $N=10$, obtained from the minimization of the gravitational potential energy 
in the linkage toy model.}
\label{fig:eq_alpha}
\end{center}
\end{figure}

\textcolor{black}{To show the generic bistability of Jacob's ladder in gravity, we here assume a uniform rotation of alternating directions, i.e., $\theta_n=\theta$.
Note that $\theta_n$ is defined as positive CCW for even-$n$ blocks and positive CW for odd-$n$ blocks, where $n=0,1,\cdots,N-1$. See Appendix A in End Matter in the main text.}
Defining $\Delta_x(\theta)$ and $\Delta_y(\theta)$ as illustrated in Fig.~\ref{fig:sm_model}, we can readily find the following trigonometric relations 
\begin{equation}
 a^2 = (s\sin\theta+\Delta_x)^2+(s\cos\theta+\Delta_y)^2,
 \quad
 \mbox{and}
 \quad
 b^2 = (s\sin\theta+\Delta_x)^2+(s\cos\theta-\Delta_y)^2,
 \label{eq:b}
\end{equation}
from which we obtain
\begin{equation}
 \Delta_x(\theta) = -s\sin\theta+\sqrt{\frac{1}{2}(a^2+b^2)-s^2\cos^2\theta-\frac{(a^2-b^2)^2}{16s^2\cos^2\theta}},
  \quad
 \mbox{and}
 \quad
 \Delta_y(\theta) = \frac{a^2-b^2}{4s\cos\theta}.
 \label{eq:del_xy}
\end{equation}
To determine the equilbrium anlge $\alpha$, we now consider a net gravitational potential $U$ of the unit strucutre shown in Fig.~2 (b) inset in the main manuscript as a function of $\theta$.
Taking the reference point of $U(\theta)$ as the geometric center of the upper block, we can write down $U(\theta)$ as
\begin{equation}
 U(\theta) = -2mg (s\sin\theta+\Delta_x) = -mg \sqrt{2(a^2+b^2)-4s^2\cos^2\theta-\frac{(a^2-b^2)^2}{4s^2\cos^2\theta}}.
 \label{eq:U_g}
\end{equation}
The plot of $U(\theta)$ given in Fig.~2 (b) reveals the symmetric double-well shape, showing that the system acquires the bistability under the action of gravity \textcolor{black}{in the linkage ($k\rightarrow\infty$) limit.} Minimization of $U(\theta)$ yields the equilibrium angles $\bar{\theta}=\pm \alpha$ given by
\begin{equation}
 \cos\alpha = \frac{\sqrt{a^2-b^2}}{2s}, 
 \quad
 \mbox{or}
 \quad
 a^2 = b^2 +4s^2\cos^2\alpha.
 \label{eq:eq_alpha}
\end{equation}
We can also obtain
\begin{equation}
 U_{\rm min} = U(\pm \alpha) = -2mgb,
 \quad
 \Delta_x(\alpha) = -s\sin\alpha+b,
 \quad
 \Delta_y(\alpha) = s\cos\alpha = \frac{1}{2}\sqrt{a^2-b^2}.
 \label{eq:min_values}
\end{equation}
Using Eq.~(\ref{eq:eq_alpha}), we can also obtain the energy barrier to overcome the unstable fixed point $\theta=0$ as
\begin{equation}
 \Delta U (\alpha) = U(0)-U(\alpha) = 2mg b\left(1-\sqrt{1-\frac{s^2}{b^2}\sin^4\alpha}\right).
 \label{eq:delta_U}
\end{equation}
For $\alpha=0$, or when $a^2=b^2+4s^2$, the bistability is lost and the ladder takes the gapless high-symmetry configuration as shown in Fig.~\ref{fig:eq_alpha}.

\section{floppy modes and index count}
Having determined the zig-zag ground state configuration specified by $\alpha$ in Eq.~(\ref{eq:eq_alpha}), we next examine the linear response of the spring system. 
To identify the zero energy modes as well as states of self-stress, we first igore the gravity but
will consider the geometric stiffness~\cite{Guest-IJSS-2005}) due to gravitional pulling later separately, which will be shown to stabilize most of the identified inifinitesimal mechanisms.
For further explanation, see SM Sec.~\ref{sec:finite_mechanisms}.

\smallskip

For the equilibrium configuration specified above, we have
\begin{equation}
 \bar{X}_{n+1}-\bar{X}_n = b,
 \quad
 \bar{Y}_{n+1}-\bar{Y}_n = \textcolor{black}{-(-1)^n} s\cos\alpha,
 \quad
 \bar{\theta}_n = \bar{\theta},
 \label{eq:eq_values}
\end{equation}
where $\bar{\theta}=\pm\alpha$. With a suitable rigid-body translation, we can set
\begin{equation}
 \bar{X}_n = nb,
 \quad
 \bar{Y}_n = (-1)^n r\cos\alpha. 
 \label{eq:Rg}
\end{equation}
where $(-1)^n$ in $\bar{Y}_n$ suggests the underlying zig-zag configuration. For convience we have defined
\begin{eqnarray}
 r &=& \frac{s}{2}.
 \label{eq:r}
\end{eqnarray} 

To investigate the linear regime, we assume the small deviations of the translational and rotational degrees of freedom from those in Eq.~(\ref{eq:Rg}) as
\begin{equation}
 X_n = \bar{X}_n +\delta u_n,
 \quad
 Y_n = \bar{Y}_n +\textcolor{black}{(-1)^n}\delta v_n,
 \quad
 \theta_n = \bar{\theta}+\delta \theta_n.
 \label{eq:small_dev}
\end{equation}
\textcolor{black}{The spring lengths, $L_{A, n}$ and $L_{B, n}$, can be written as
\begin{eqnarray}
 L_{A,n}^2 &=& \left[b+(\delta u_{n+1}-\delta u_n)-r(\sin\theta_{n+1}-\sin\theta_{n})\right]^2+\left[s\cos\alpha+(\delta v_{n+1}+\delta v_n)+r(\cos\theta_{n+1}+\cos\theta_{n})\right]^2, 
 \label{eq:ell_A}\\
 L_{B,n}^2 &=& \left[b+(\delta u_{n+1}-\delta u_n)+r(\sin\theta_{n+1}-\sin\theta_{n})\right]^2+\left[s\cos\alpha+(\delta v_{n+1}+\delta v_n)-r(\cos\theta_{n+1}+\cos\theta_{n})\right]^2.
 \label{eq:ell_B}
\end{eqnarray}
}
At the mechanical equilibrium, we have $L_{A, n}=a$ and $L_{B, n}=b$.
Plugging Eqs.~(\ref{eq:small_dev}) into Eqs.~(\ref{eq:ell_A}) and (\ref{eq:ell_B}), and retaining the terms up to the first order of the small quantites, 
$\delta u, \delta v$ and $\delta\theta$, we find
\begin{align}
 L_{A, n}^2 &\simeq a^2 + 2b (\delta u_{n+1}-\delta u_n)\textcolor{black}{+4s\cos\alpha (\delta v_{n+1}+\delta v_n)}
 	-\left[s\cos\alpha(b+2s\sin\bar{\theta})\right]\delta\theta_{n+1}+\left[s\cos\alpha(b-2s\sin\bar{\theta})\right]\delta\theta_{n},
\label{eq:ell_A2}\\
L_{B, n}^2 &\simeq b^2 + 2b (\delta u_{n+1}-\delta u_n) + (sb\cos\alpha)(\delta\theta_{n+1}-\delta\theta_{n}).
\label{eq:ell_B2}
\end{align}
\textcolor{black}{In general,} the extension $e_n$ of a spring of natural length $\bar{L}$ is related to its actual length $L_n$ as 
\begin{equation}
 e_n  = L_n-\bar{L} = \frac{L_n^2-\bar{L}^2}{L_n+\bar{L}} \simeq \frac{L_n^2-\bar{L}^2}{2\bar{L}},
 \label{eq:e_n}
\end{equation}
which is valid when $e_n/\bar{L} \ll 1$.
Substituting Eqs.~(\ref{eq:ell_A}) and (\ref{eq:ell_B}) into Eq.~(\ref{eq:e_n}), we obtain
\begin{align}
 e_{A, n} &\simeq \frac{b}{a}(\delta u_{n+1}-\delta u_n)\textcolor{black}{+4\gamma (\delta v_{n+1}+\delta v_n)}-c_1 \delta \theta_{n+1}+c_2\delta \theta_n,
 \label{eq:eA}\\
 e_{B, n} &\simeq \delta u_{n+1}-\delta u_n +\gamma a (\delta \theta_{n+1}-\delta \theta_n),
 \label{eq:eB}
\end{align}
where we have defined
\begin{equation}
 \gamma = \frac{s}{2a}\cos\alpha = \frac{1}{4}\sqrt{1-\frac{b^2}{a^2}},
 \quad
 c_1 = \gamma (b+2s\sin\bar{\theta}),
 \quad
 c_2 = \gamma (b-2s\sin\bar{\theta}).
 \label{eq:c_12}
\end{equation}
Because we are considering a periodic lattice system, we move to the Bloch space by fourier-transforming the all relevant physical qunatities
and work with the wavenumbers $q$ in the 1st Brillouin zone of $-\pi/b \leq q\leq \pi/b$:
\begin{equation}
  \delta u_n = \sum_{q} \hat{u}(q)e^{iqnb},
  \quad
  \delta v_n = \sum_{q} \hat{v}(q)e^{iqnb}
  \quad
  \delta\theta_n = \sum_{q} \hat{\theta}(q) e^{iqnb},
  \quad
   e_{A(B), n} = \sum_{q} \hat e_{A (B)}(q)e^{iqnb},
  \label{eq:FT}
\end{equation}
where we have used the fact that the effective lattice constant (along $x$-axis) is $\bar{X}_{n+1}-\bar{X}_n = b$ from Eq.~(\ref{eq:eq_values}).
Plugging Eqs.~(\ref{eq:FT}) into Eqs.~(\ref{eq:eA}) and (\ref{eq:eB}), we obtain the linear relation between
the spring extension and the node displacements as 
\begin{eqnarray}
\left(
\begin{array}{c}
  \hat{e}_A (q) \\
  \hat{e}_B(q) 
\end{array}
\right)
 &=& 
 {\bf C}(q) 
\left(
\begin{array}{c}
  \hat{u}(q) \\
  \hat{v}(q)\\
  \hat{\theta}(q) 
\end{array}
\right),
\end{eqnarray}
where the compatibility matrix in the Bloch space ${\bf C}(q)$ is the $2 \times 3$ matrix given by
\begin{eqnarray}
 {\bf C}(q) &=&
\left(
\begin{array}{ccc}
  (b/a)(e^{iqb}-1) &  \textcolor{black}{4\gamma (e^{iqb}+1)}  & -c_1e^{iqb}+c_2 \\
  e^{iqb}-1 &  0 &  \gamma a (e^{iqb}-1) 
\end{array}
\right).
\label{eq:C-q}
\end{eqnarray}
This is a straightforward extension of the compatibility matrix of the Kane-Lubensky chain by including the translational degrees of freedom of the constituting rotors.

\smallskip

\begin{figure}[htbp]
\begin{center}
 \includegraphics[width=0.70\linewidth]{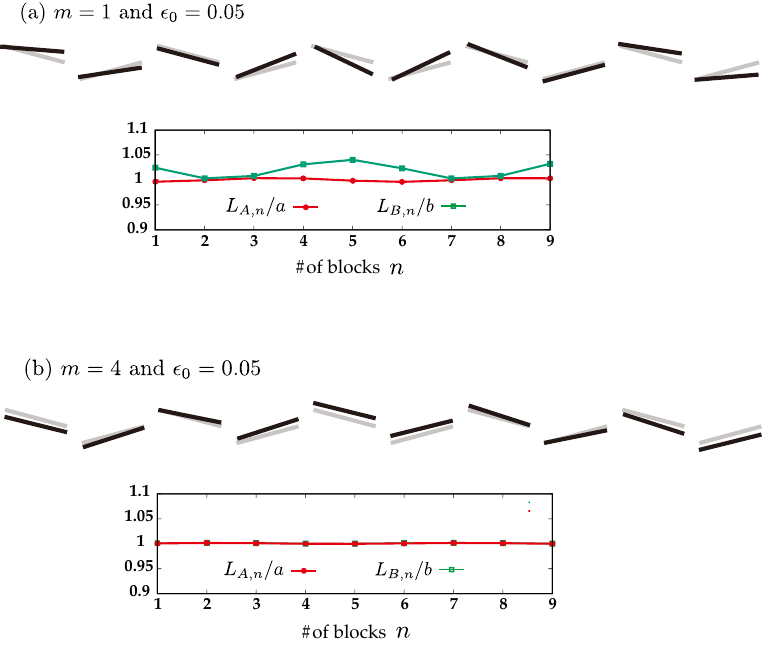}
\caption{Plots of the deformed ZM configurations predicted from Eq.~(\ref{eq:zm_n}) \textcolor{black}{and Eq.~(\ref{eq:small_dev})} for $N=10$ ladder. The black bars represent the deformed ZMs, and the gray ones represent the equilibirum configuration given in Eq.~(\ref{eq:eq_values}). 
\textcolor{black}{Relative spring extensions, $L_{A,n}/a$ and $L_{B,n}/b$, calculated from the predicted ZMs in Eq.~(\ref{eq:zm_n}) and Eq.~(\ref{eq:small_dev}), are also
shown. } (a) $p=1$ and $\epsilon_0=\textcolor{black}{0.05}$ and (b) $p=\textcolor{black}{4\, (=N/2-1)}$ and $\epsilon_0=0.05$.
The other parameters are $b/s=1.2$ and $\alpha=75^{\circ}$, where $a=\sqrt{b^2+4s^2\cos^2\alpha}$ in (a) and (b). }
\label{fig:zm_plot}
\end{center}
\end{figure}

It is easy to find the ZMs that are the nullspace vectors of ${\bf C}(q)$ given by
\begin{equation}
 {\bf U}_0(q) = 
 \left(
\begin{array}{c}
  \hat{u}_0(q) \\
  \hat{v}_0(q)\\
  \hat{\theta}_0(q) 
\end{array}
\right)
=
 \epsilon(q) \left(
\begin{array}{c}
  -2\gamma a (e^{iqb}\textcolor{black}{+}1) \\
  \textcolor{black}{(b+s\sin\bar{\theta})e^{iqb}-(b-s\sin\bar{\theta})}\\
  2(e^{iqb}\textcolor{black}{+}1)  
\end{array}
\right)
\label{eq:zm_q}
\end{equation}
where $\epsilon(q)$ is an arbitrary $q$-dependent parameter, albeit inifinitesimally small.
Equation~(\ref{eq:zm_q}) ensures $n_0(q)=1$ \textcolor{black}{for $q>0$.} 
\textcolor{black}{For $q=0$, we find two ZMs, which are given by
${\bf U}_0(q=0)=(u,0,0)$ and $U_0(q=0) = \epsilon(q=0)(0, 2s\sin\alpha, 4)$.
The former represents a rigid-body translation along $x$ and the latter comes from Eq.~(\ref{eq:zm_q}).
(In fact, the $x$-component of the latter can be arbitrary and is set zero here.)} 
Thus we confirm $N_0=\sum_{q} n_0(q) =N+1$-independent ZMs.
For example, by choosing $\epsilon(q)= (\epsilon_0/2) (\delta_{q, k}+\delta_{-q,k})$, where $\delta_{\alpha\beta}$ is the Kronecher delta and $\epsilon_0$ is 
some small parameter, we obtain in the real space 
a set of the compact anaytical expressions of the ZMs given by
\begin{equation}
  {\bf U}^{(0)}_n = 
 \left(
\begin{array}{c}
  \delta u^{(0)}_n \\
  \delta v^{(0)}_n\\
  \delta\theta^{(0)}_n 
\end{array}
\right)
=
 \epsilon_0 \left(
\begin{array}{c}
  -2\gamma a \left[\cos((n+1)kb)\textcolor{black}{+}\cos(nkb)\right] \\
  \textcolor{black}{-}(b-s\sin\bar{\theta})\cos(nkb)\textcolor{black}{+}(b+s\sin\bar{\theta})\cos((n+1)kb) \\
    2\left[\cos((n+1)kb)\textcolor{black}{+}\cos(nkb)\right]
\end{array}
\right).
\label{eq:zm_n}
\end{equation}
In Fig.~\ref{fig:zm_plot}, we show the deformed configurations of the $N=10$ ladder predicted from Eq.~(\ref{eq:zm_n}) for the mode numbers $p=1$ and \textcolor{black}{4}, where
the wavenumber is $k=2\pi p/(Nb)$, and the amplitudes $\epsilon_0=\textcolor{black}{0.05}$.

\smallskip

The equilibrium matrix, ${\bf Q}(q)$, defined through the force balance condition, is known from ${\bf Q}(q)={\bf C}^{\dagger}(q)$, which is thus $3\times 2$ matrix given by
\begin{equation}
\left(
\begin{array}{c}
  \hat{f}_x (q) \\
  \hat{f}_y (q) \\
  \hat{m}(q) 
\end{array}
\right)
 =
 {\bf Q}(q) 
\left(
\begin{array}{c}
  \hat{t}_A(q) \\
  \hat{t}_B(q)
 \end{array}
  \right),
  \quad
  {\bf Q}(q) =
\left(
\begin{array}{ccc}
  (b/a)(e^{-iqb}-1) &  e^{-iqb}-1 \\
  \textcolor{black}{4\gamma (e^{-iqb}+1)}  & 0 \\
  -c_1e^{-iqb}+c_2 & \gamma a (e^{-iqb}-1) 
\end{array}
\right).
\label{eq:Q}
\end{equation}
where $(t_A, t_B)$ are the tensions in the springs A and B, and $(f_x,f_y)$ and $m$ represent the net forces and moment (about the center of mass) on each block, respectively.
(The quantities appearing in Eq.~(\ref{eq:Q}) are the Fourier components of them.)
Clearly, there is no SSS for \textcolor{black}{$q\neq 0$ when $a/b>1$.
For $q=0$, however, there is one SSS given by $t_A=0$ and $t_B=t$, which represents a system-spanning SSS along the green links lined up straight. See Fig.~2 (c).
We thus find $N_S=\sum_{q}n_s(q)=1$. }
We confirm $N_0-N_S=N$, as the Index theorem requires.

\textcolor{black}{\section{Finite mechanisms}~\label{sec:finite_mechanisms}
In the absence of gravity, all of the $N$ inpendent ZMs are likely to be finite mechanisms that satisfy
\begin{equation}
L_{A,n} = a, 
\quad
\mbox{and}
\quad
L_{B,n} = b,
\label{eq:zm-eqs}
\end{equation}
where $L_{A,n}$ and $L_{B,n}$ are given in Eqs.~(\ref{eq:ell_A}) and (\ref{eq:ell_B}), respectively.
}

\begin{figure}[htbp]
\begin{center}
 \includegraphics[width=0.89\linewidth]{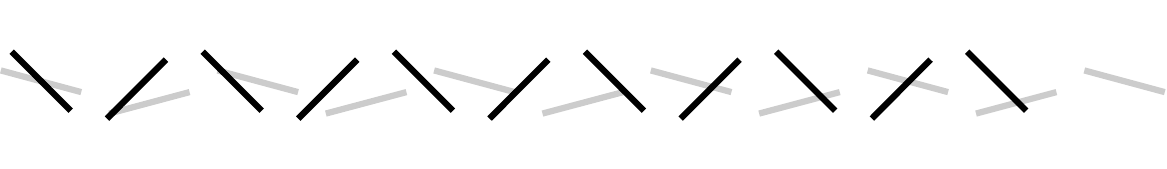}
\caption{Plot of a uniform finite mechanism predicted from Eqs.~(\ref{eq:u-zms}) and (\ref{eq:v-zms}) for $\theta=45^{\circ}$ (solid black bars). 
The chain consists of $N=10$ blocks with an equilibrium zig-zag angle of $\alpha=75^{\circ}$ and $b/s=1.2$ (solid gray bars).}
\label{fig:finite_zm}
\end{center}
\end{figure}

\textcolor{black}{
For a uniform (zig-zag) rotation given by $\theta_n=\theta =$ (const) for all $n$,  it is easy to find analytical solutions of Eqs.~(\ref{eq:zm-eqs}) in terms of $(\delta u_n, \delta v_n)$, given by
\begin{equation}
 \delta u_{n+1} -\delta u_n =-b+\sqrt{\frac{a^2+b^2}{2}-s^2\left(\cos^2\theta+\frac{\cos^4\alpha}{\cos^2\theta}\right)}, 
 \label{eq:u-zms}
\end{equation}
and
\begin{equation}
 \delta v_{n+1}+\delta v_n =  s\left(\frac{\cos\alpha}{\cos\theta}-1\right)\cos\alpha.
 \label{eq:v-zms}
\end{equation}
It is also easy to see the following:
\begin{eqnarray}
 \delta u_{n+1}-\delta u_n &=& \Delta_x(\theta)-\Delta_x(\alpha) +s(\sin\theta -\sin\alpha), \\
 \delta v_{n+1}+\delta v_n &=& \Delta_y(\theta)-\Delta_y(\alpha),
 \label{eq:deltas}
\end{eqnarray}
where $\Delta_x(\theta)$ and $\Delta_y(\theta)$ have been defined in Eqs.~(\ref{eq:del_xy}).
For example, we show in Fig.~\ref{fig:finite_zm} the configuration predicted from Eqs.~(\ref{eq:u-zms}) and (\ref{eq:v-zms}) for $\theta=45^{\circ}$, for the $N=10$ ladder of its ground state configuration with $\alpha=75^{\circ}$.
}

\textcolor{black}{
While an explicit calculation will be quite involved (and may not be so insightful in any case), in general, there may be $N$ independent solutions $(\delta u_n,\delta v_n,\delta\theta_n)$ satisfying Eqs.~(\ref{eq:zm-eqs}), constituting the $N$-independent finite mechanisms.
}

\textcolor{black}{
In our system, the spring elastic energy, $E_{\rm el}$, itself does not a priori provide a bistable energy landscape, unlike the 1 DoF chain model~\cite{Chen-PNAS-2014}. 
However, when the ladder is subjected to a tension due to the gravity, many of those finite mechanisms may disappear.
The graivty selects the two energetically degenerated ground states out from many otherwise equal zero-energy configurations. 
This is the first effect of the gravitational pulling.
However, the magnitude of strains in springs due to the gravitational pulling is $\sim mg/ks$, which are assumed sufficiently small and such "pre-strains" will be entirely nelgected in the following analysis.
Yet, this pre-stretch provides the chain some geometric sitffness of rigidities $\sim mg/b$, which is the second effect of the gravity that we take into account.
Some of the bulk ZMs suppressed by the pre-tension may be still considered as soft modes as their energay gap goes to zero for associated wave number $q\rightarrow 0$. 
Many others are localized modes of $q\sim 2\pi/b$, which have higher energies and are well separated from the ZM localized at the chain edge.
Neglecting those bulk ZMs, we can effectively reduce the number of DoFs in the problem; at this point, the elastic energy $E_{\rm el}$ finds its bistable energy landscape.
Additionally, we also need to consider the rotational dynamics of the blocks driven by the net gravitational torques, as we will do so for the continuum description of the kink propagation.
This is the third effect of the gravity.
}

\smallskip

\textcolor{black}{
While a direct evaluation of the total potential energy (spring+gravity) cannot be implemented due to the difficulty of analytical treatment, our balanced approach outlined above 
will be shown to work well for a sufficiently stiff (i.e., a near-linkage) system such as $mg/ks \ll1$ but still the stiffness $\sim mg/b$, is large enough to effectively suppress the bulk ZMs inherent in the original ladder.
}

\section{Reduction to Kane-Lubensky chain model}
As expected from Fig.~2 (c) in the main manuscript, our model can be reduced to the KL chain model by fixing, for example, the positions of the nodes of the blocks that are connected to 
the springs B. 
The node of the $n$-th block can be written as
\begin{eqnarray}
 {\bm x}_{B, n} &=& \left(\bar{X}_n+\delta u_n +r\sin\textcolor{black}{\theta_n}, \, \, \bar{Y}_n+(-1)^n(\delta v_n -r \cos \theta_n)\right),
 \label{eq:R_B}
\end{eqnarray}
where we have assumed Eq.~(\ref{eq:Rg}) here, i.e., $\bar{X}_n=nb$ and $\bar{Y}_n=(-1)^n r\cos\alpha$.
Thus, if we impose 
\begin{equation}
 \delta u_n = -r(\sin\theta_n-\sin\bar{\theta}), 
 \qquad
 \delta v_n = r(\cos\theta_n-\cos\alpha),
 \label{eq:KL-constraints}
\end{equation}
for $n$, we exactly have ${\bm x}_{B, n}=(nb+r\sin\bar{\theta},0)$, and obviously the ladder becomes the KL chain.

\smallskip

At the first order in the small quantities, Eq.~(\ref{eq:KL-constraints}) requires
\begin{equation}
 \delta u_n = -r\cos\bar{\theta} \delta \theta_n,
 \qquad
 \delta v_n = -r\sin\bar{\theta}\delta \theta_n.
 \label{eq:u-v-theta}
\end{equation}
Plugging these into Eqs.~(\ref{eq:eA}) and (\ref{eq:eB}), and using Eqs.~(\ref{eq:c_12}), we immediately obtain
\begin{equation}
 e_{n,A} = -2c_1\delta\theta_{n+1}+2c_2\delta\theta_n,
 \qquad
 e_{n,B} = 0.
 \label{eq:e-KL}
\end{equation}
This is essentially the same linear relation studied in the KL chain model.
Note the difference in the factor 2 in the definition of $c_{1,2}$ between Ref.~\cite{Kane-NatPhys-2014, Lubensky-Review-2015} and ours.

\section{Topological winding number}
Here we discucss with an integer winding nubmer $n_C$ of the equilibrium matrix ${\bf C}$ in the Brillouin space.
A non-trivial ZM exists only when the determinant $\det|{\bf C}(q)| = C(q) e^{i\phi(q)}$ encloses the zero in its complex plane when $q$ moves through the
1st Brillouin zone. 
The curve is a circle of radius $c_1$ and center $(-c_2,0)$, and its topology is characterized by the integer winding number $n_C$ given by 
\begin{equation}
 n_C = \frac{1}{2\pi}\int_0^{2\pi} d\phi(q)  = \frac{1}{2\pi i}\int_0^{2\pi/b} dq \frac{d}{dq} \ln \det |{\bf C}(q)|.
 \label{eq:n}
\end{equation}
For the square matrix  ${\bf C}(q)$ given in Eq.~(4) in the main text, we have
\begin{eqnarray}
 \det|{\bf C}(q)| &=& 2(e^{iqb}-1)\left(c_1e^{iqb}-c_2\right).
 \label{eq:det-C} 
\end{eqnarray}
Plugging this into Eq.~(\ref{eq:n}) with the variable change $z=e^{iqb}$, we find
\begin{eqnarray}
  n_C &=& \frac{1}{2\pi i} \oint_{C} \frac{dz}{z-c_2/c_1}+\frac{1}{2\pi i} \oint_{C} \frac{dz}{z-1},
  \label{eq:n-2}
\end{eqnarray}
where $C$ is the circular path of unit radius enclosing the origin in the $z$ plane.
Because we simply exclude $q=0$ (rigid-body translation), the second term in Eq.~(\ref{eq:n-2}) gives always zero, which leads to 
\begin{equation}
 n_C = 1
 \quad
 \mbox{for}
 \quad
 0 < \bar{\theta} < \pi/2,
 \quad
 {\rm and}
 \quad
 n_C = 0 
 \quad
 \mbox{for}
 \quad
 -\pi/2 < \bar{\theta} < 0.
 \label{eq:winding}
\end{equation}
The topolgoical analysis thus suggests that the ZM localizes at the top for $\bar{\theta}=\alpha$ configuration [Fig.~2 (c) top] and at the bottom for $\bar{\theta}=-\alpha$ configuration [Fig.~2 (c) bottom], where where we assumed $0 <\alpha<\pi/2$ here. \\

\textcolor{black}{
For completeness, we here outline an explicit, direct calculation of the winding number given in Eq.~(\ref{eq:n}).
First, in order to avoid any pole just on the integration path of unit circle on the complex $z$ plane, we may consider
\begin{eqnarray}
 \det |{\bf C}(q)| &=& 2(z-a)(c_1z-c_2),
 \label{eq:C-a}
\end{eqnarray}
where we set $a=1+\epsilon$ and we will take $\epsilon \rightarrow +0$ later on.
We also write $k=qb$ in the following. The phase $\phi(k)$ is explicitly given from Eq.~(\ref{eq:C-a}) by
\begin{eqnarray}
 \tan\phi(k) &=& \frac{c_1\sin(2k)-(ac_1+c_2)\sin k}{c_1\cos(2k)-(ac_1+c_2)\cos k +ac_2}.
\end{eqnarray}
After a long but straightforward calcuation, we may obtain
\begin{eqnarray}
\frac{d\phi(k)}{dk} &=& 1+\frac{1}{2}\left[\frac{1-a^2}{1+a^2-2a\cos k}+\frac{c_1^2-c_2^2}{c_1^2+c_2^2-2c_1c_2 \cos k}\right].
\end{eqnarray}
Integrating this over the 1st Brillouin zone, $0\leq k\leq 2\pi$, we have 
\begin{eqnarray}
 n_C &=& 1+\frac{1}{2}\left(\frac{1-a^2}{|1-a^2|}+\frac{c_1^2-c_2^2}{|c_1^2-c_2^2|}\right).
\end{eqnarray}
Setting $a=1+\epsilon$ with $\epsilon\rightarrow +0$, we obtain
\begin{eqnarray}
 n_C &=& \frac{1}{2}\left(1+\frac{c_1^2-c_2^2}{|c_1^2-c_2^2|}\right),
\end{eqnarray}
which gives the correct winding number as given in Eq.~(\ref{eq:winding}).
Note that setting $a=1$ from the beginning may lead to an aritificial $+1/2$ contribution to $n_C$.}

Additionally, for $a=b$, the curve contracts towards the origin \textcolor{black}{on the complex $({\rm Re}\det{\bf C}, {\rm Im}\det{\bf C})$ plane}, making its topological phase undefined at the linear order. 

\section{phonon dispersion}
When all the A (red) and B (green) springs have the spring constant $k$, the dynamical matrix ${\bf D}(q)$ becomes
\begin{eqnarray}
 {\bf D} &=& 
 k {\bf Q}  {\bf Q}^{\dagger},
  \label{eq:D}
\end{eqnarray}
which is the $3\times3$ square matrix.
To make the following argument analytically tractable, we continue to work with the additional constraints given in the main manuscript, \textcolor{black}{$v_n=-r\sin\bar{\theta} \delta\theta_n$}, 
for which ${\bf D}(q)$ is reduced to the $2\times2$ matrix. The equations of motion for the Fourier mode $q$ are then given by
\begin{eqnarray}
\left(
\begin{array}{cc}
  \textcolor{black}{2m} & 0 \\
  0 & I 
\end{array}
\right)
\frac{d^2}{dt^2}\left(
\begin{array}{c}
  \hat{u}(q) \\
  \hat{\theta}(q)
\end{array}
\right)
&=& -{\bf D}(q) 
\left(
\begin{array}{c}
  \hat{u}(q) \\
  \hat{\theta}(q)
\end{array}
\right)
\label{eq:D}
\end{eqnarray}
where \textcolor{black}{$I=ms^2/2$} is the moment of inertia of a block about its center. 

\smallskip

\begin{figure}[htbp]
\begin{center}
 \includegraphics[width=0.99\linewidth]{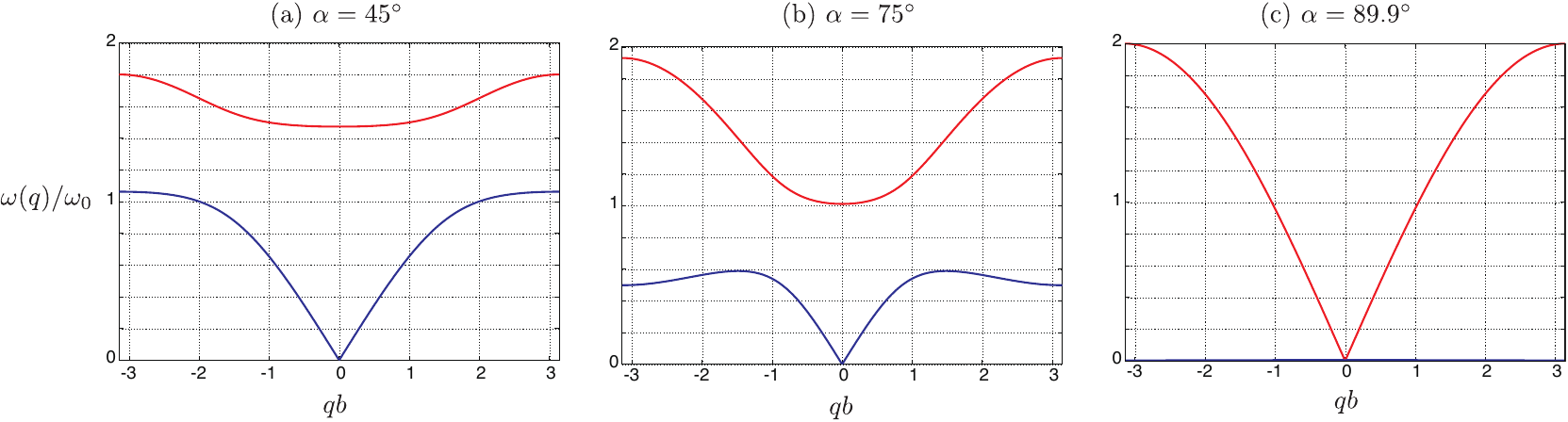}
\caption{Plots of $\omega(q)/\omega_0$ as a function of $qb$, Eq.~(\ref{eq:dispersion}), for different values of $\alpha$ shown with $b/s=1.3$. 
Red and blue lines represent $\omega_{+}(q)$ and $\omega_{-}(q)$, respectively.}
\label{fig:dispersion}
\end{center}
\end{figure}

A straightforward calculation yields the explicit analytical expressions of ${\bf D}(q)$ given by
\begin{eqnarray}
 D_{11}/k &=& 2\left(1+\frac{b^2}{a^2}\right)(1-\cos(qb)), \\
 D_{12}/k &=& 2\gamma \left(a-\frac{b^2}{a}\right)(1-\cos(qb))+2i b\frac{s^2}{a^2}\sin(2\alpha)\sin(qb), \\
 D_{22}/k &=& (v_1-v_2)^2+2\left[v_1v_2+(\gamma a)^2\right](1-\cos(qb)), 
 \label{eq:D_comp}
\end{eqnarray}
and $D_{21} =D^{\ast}_{12}$, \textcolor{black}{where $v_1 = c_1+2\gamma s \sin\bar{\theta}$ and $v_2=c_2-2\gamma s \sin\bar{\theta}$, as given in the main text.}
Applying the standard eigenvalue analysis to this, we obtain the dispersion relations given by
\begin{eqnarray}
 \omega_{\pm}^2(q) &=& \frac{\omega_0^2}{\textcolor{black}{4}} \left[d_{11}+d_{22} \pm\sqrt{(d_{11}-d_{22})^2+4 d_{12}^2}\right],
 \label{eq:dispersion}
\end{eqnarray}
where \textcolor{black}{$\omega_0=\sqrt{k/m}$} and 
\begin{align}
 d_{11}(q) &= \left(1+\frac{b^2}{a^2}\right) (1-\cos(qb)), \\
 d_{22}(q) &= \frac{8s^2}{a^2}\sin^2(2\alpha)+\left[\left(1+\frac{b^2}{a^2}\right)\cos^2\alpha-\frac{4s^2}{a^2}\sin^2(2\alpha)\right](1-\cos(qb)), \\
 d^2_{12}(q) &= \left(1-\frac{b^2}{a^2}\right)^2\cos^2\alpha (1-\cos(qb))^2+\frac{4b^2s^2}{a^4}\sin^2(2\alpha)\sin^2(qb).
 \label{eq:d_s}
\end{align}
Specifically, we find
\begin{equation}
 \omega_+(q=0) = \textcolor{black}{2\omega_0} \frac{s}{a}\sin(2\alpha),
 \quad
 \omega_-(q=0) = 0.
 \label{eq:gap}
\end{equation}
Thus, for general $\alpha$ except $\alpha=0$ or $90^{\circ}$, there is the energy gap at $q=0$ in $\omega_+$, which is related to the rotational modes of the blocks.
On the other hand, the translation of the blocks shows the acoustic mode characterized by $\omega_-(q\rightarrow 0)=0$.
In Fig.~\ref{fig:dispersion}, $\omega_{\pm}(q)$ is plotted as a function of $qb$ for several representative values of $\alpha$.

\section{Details of numerical simulation}~\label{sec:simulation}
In our numerical model, the configuration of the $n$-th block is specified by the center-of-mass position ${\bm X}_n=(X_n,Y_n)$ and the rotational angle measured from the veritcal axis, $\theta_n$.
The positions of A and B nodes in the $n$-th block are then given by
\begin{equation}
 {\bm x}_{A,n} = {\bm X}_n-r{\bm t}_n,
 \quad
 {\bm x}_{B,n} = {\bm X}_n+r{\bm t}_n, 
 \label{eq:x_A-x_B}
\end{equation}
where $r=s/2$ [Eq.~(\ref{eq:r})] and its orientation ${\bm t}_n = \sin\theta_n \hat{\bm e}_x-(-1)^n\cos\theta_n\hat{\bm e}_y$.
Writing the spring lengths $L_{\sigma,n} = |{\bm x}_{\sigma, n+1}-{\bm x}_{\sigma, n}|$, where $\sigma=$ A or B, the total potential energy of our system, $E$, is given by
\begin{eqnarray}
 E &=& \sum_{n=0}^{N-1} \frac{k}{2}\left[(L_{A, n}-a)^2+(L_{B, n}-b)^2\right] -\textcolor{black}{2mg^{\ast}} X_{n},
 \label{eq:E-sim}
\end{eqnarray}
where $m$ is the mass equal to A and B nodes. The total mass of one block is thus $2m$.

The dynamics of ${\bm X}_n$ and $\theta_n$ follows the Newton's equations of motion given by 
\begin{equation}
 \textcolor{black}{2m} \ddot{\bm X}_n = -\frac{\partial E}{\partial{\bm X}_n}-\zeta \dot{\bm X}_n,
 \quad
 I_c \ddot{\theta}_n = -\frac{\partial E}{\partial{\theta}_n}-\zeta  r^2 \dot{\theta}_n,
 \label{eq:eqm-sim}
\end{equation}
where we have introduced the translational (and rotational) damping parameter, $\zeta$, to ensure the steady state propagation of kinks in the chain.
The moment of inertia of a block about its center ${\bm X}_n$ is $I_c=2mr^2 = ms^2/2$.\\

\textcolor{black}{
To obtain the force and torque at each node, we now consider a variation of the energy $E$, which is most conveniently written in terms of $({\bm x}_A, {\bm x}_B)$.
From Eq.~(\ref{eq:x_A-x_B}), together with the variational relation
\begin{eqnarray}
 \delta {\bm t}_n &=& (\sigma_n \delta \theta_n \hat{\bm e}_z) \times {\bm t}_n,
 \label{eq:del_t_n}
\end{eqnarray}
we have 
\begin{equation}
 \delta {\bm x}_{A, n} = \delta {\bm X}_n-r(\sigma_n \delta \theta_n \hat{\bm e}_z) \times {\bm t}_n,
 \quad
 \delta {\bm x}_{B, n} = \delta {\bm X}_n+r(\sigma_n \delta \theta_n \hat{\bm e}_z) \times {\bm t}_n,
 \label{eq:del_x}
\end{equation}
where we have defined $\sigma_n=(-1)^n$ here.
Using Eqs.~(\ref{eq:del_x}), we obtain
\begin{eqnarray}
 \delta E &=& -{\bm f}_{A,n}\cdot \delta {\bm x}_{A, n}-{\bm f}_{B,n}\cdot\delta {\bm x}_{B, n}
  = -({\bm f}_{A, n}+{\bm f}_{B, n})\cdot \delta {\bm X}_n -\sigma_n r\left[{\bm t}_n\times ({\bm f}_{B, n}-{\bm f}_{A, n})\right]\cdot \hat{\bm e}_z \delta\theta_n,
\end{eqnarray}
which yields
\begin{eqnarray}
 {\bm f}_n &=& -\frac{\partial E}{\partial {\bm X}_n} = {\bm f}_{A, n}+{\bm f}_{B, n},\\
 T_n &=& -\frac{\partial E}{\partial \theta_n} = \sigma_n r\left[{\bm t}_n\times ({\bm f}_{B, n}-{\bm f}_{A, n})\right]\cdot \hat{\bm e}_z.
 \label{eq:f-and-T}
\end{eqnarray}
Here, it is straightforward to find
\begin{eqnarray}
 {\bm f}_{\sigma, n} &=& -\frac{\partial E}{\partial {\bm x}_{\sigma, n}} = k \left(1-\frac{\bar{L}_{\sigma}}{L_{\sigma, n}}\right) ({\bm x}_{\sigma, n+1}-{\bm x}_{\sigma, n})
 	-k \left(1-\frac{\bar{L}_{\sigma}}{L_{\sigma, n-1}}\right) ({\bm x}_{\sigma, n}-{\bm x}_{\sigma, n-1}) - mg^{\ast} \hat{\bm e}_x,
\end{eqnarray}
for $\sigma=A, B$ and we have defined $\bar{L}_A=a$ and $\bar{L}_B=b$. \\
}

We mainly simulate a relatively short chain consisting of $N=10$ blocks corresponding to our experiments, but longer chains of $N=20-50$ are also investigated. 
Choosing the block length $s=2r$, the inverse of the natural frequency $\tau=\sqrt{m/k}$, and $ks$ as the units of length, time and force, we rescale the equations of motion given above appropriately.
We set $b/s=1.3$, and $a$ is determined from the geometric relation, Eq.~(\ref{eq:eq_alpha}), for a given $\alpha$.
We also set the values of the rescaled linear damping coefficient $\zeta \tau/2m=0.005$ and the rescaled gravitational force $G=g^{\ast}\tau^2/s=mg^{\ast}/ks=0.0002$. \\

\textcolor{black}{
We used the 2nd order Runge-Kutta method to numerically integrate the rescaled dynamical equations with the non-dimensional time steps typically $0.002-0.005$ that ensures a sufficient numerical accuracy.
Total simulation time steps are typically $2-5\times 10^6$, with the active rotation of the first block of the angle $2\alpha$ completed typically within the $10^3$ non-dimensional initial time.
For kink propagation, the first block ($n=0$) is actively rotated with respect to ${\bm x}_{B,0}$  that is fixed,  while for antikink propagation, the first block is rotated with respect to the fixed ${\bm x}_{A,0}$. \\
}
\textcolor{black}{
By running simulations with varying time steps $0.0004-0.01$, we have confirmed the convergence of our numerical scheme.
Since the dynamics is essentially overdamped due to friction, the numerical scheme is highly stable.\\
}

\begin{figure}[htbp]
\begin{center}
 \includegraphics[width=0.80\linewidth]{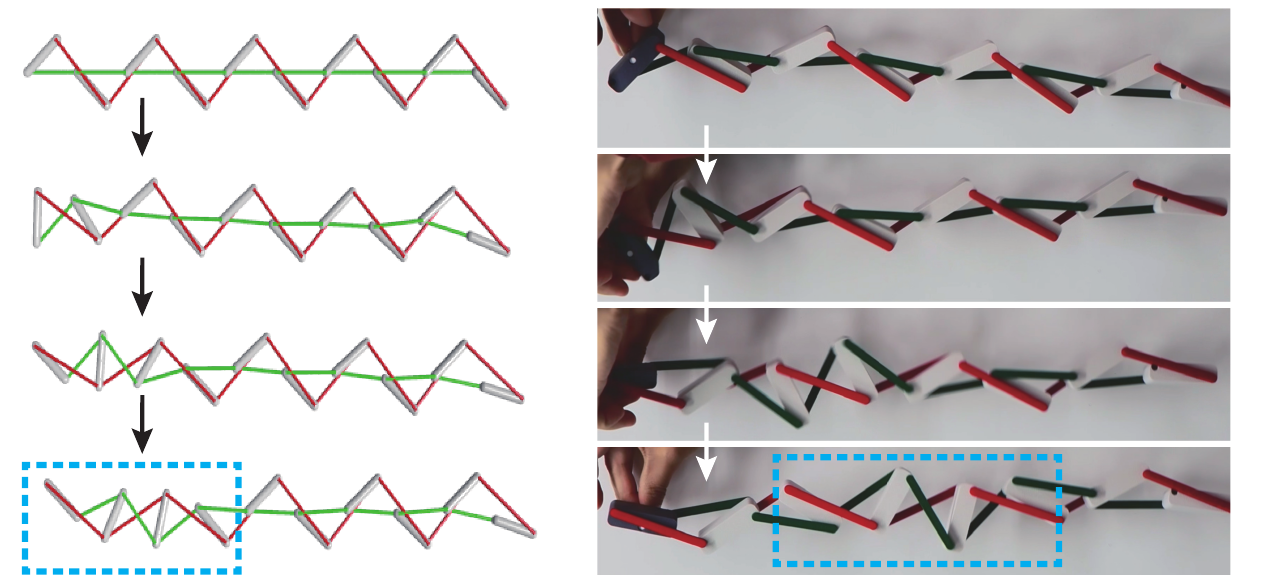}
\caption{\textcolor{black}{(Left) Numerically observed antikink behavior for $N=10$ and $\alpha=50^{\circ}$. Large amplitude domain wall created at the left edge moves in the bulk
but soon stops and stays there, with a self-stress like configuration. (Right) The same behavior is observed in simple experiment using a 3D printed linkage model.
The antikink also stops in the middle. See also SM Movie 10.
The region enclosed by the blue dashed line in the simulation (left) and the physical linkage model (right) are observed to take the same local configuration.}}
\label{fig:antikink-50}
\end{center}
\end{figure}

\textcolor{black}{
Typical simulation results for the different equilbrium angles, $\alpha=50^{\circ}, 60^{\circ}$ and $87^{\circ}$, are given in Supplemental Movies 4-9.
We find that antikink propagetes from left to right for $\alpha =60^{\circ}$ and 87$^{\circ}$ (Supplemental Movies 5 and 7).
In contrast, for $\alpha=50^{\circ}$ (and for any $\alpha$ less than this), anti-kink stops halfway and remains in that position, with a self-stress like configuration; see Fig.~\ref{fig:antikink-50}.
Since there are two localized ZMs (one at each edge)  in this configuration, one localized SSS is expected at the domain boundary from the index theorem.
Still, we do not yet completely understand when this domain wall propagates and when not, which we hope to clarify in our future publications.
}

\textcolor{black}{
\section{Continuum description}
In this section, we develop a continuum theory of our asymmetric Jacob's ladder. 
Our starting point is the elastic energy of the spring system given by
\begin{eqnarray}
 E_{\rm el} &=& \sum_{n=0}^{N-1} \frac{k}{2}(e_{A,n}^2+e_{B,n}^2).
 \label{eq:E-spring}
\end{eqnarray}
The effects of gravity will be suitably taken into account according to the protocol outlined in SM Sec.~\ref{sec:finite_mechanisms}. 
In this approach, we assume that the transverse fluctuations $\delta v$ are effectively suppressed due to the pretension by gravitational pulling.
In fact, for a kink propagation, the green linkages are observed to keep its almost straight configuration, while the red ones rotate (see Fig.~3 (a) in the main text and Supplemental Movie 4).
The validity of our assumption will then be examined by comparing the analytical predictions with our numerical simulation that precisely includes all the effects of gravity (See Fig.~3 (d) and SM Sec.~\ref{sec:simulation}).
}

\textcolor{black}{
\subsection{Kink propagation}
For a kink, we assume $\delta v_n = r(\cos\theta_n-\cos\alpha)$, so that $y_{B,n}=0$ as given in Eq.~(\ref{eq:KL-constraints}), while $u_n$ are still treated as the independent variable.
We then find
\begin{equation}
 L_{A,n}^2 = \left[b(1+\eta_n)-r(\sin\theta_{n+1}-\sin\theta_{n})\right]^2 + \left[2r(\cos\theta_{n+1}+\cos\theta_{n})\right]^2, 
\ 
 L_{B,n}^2 = \left[b(1+\eta_n)+r(\sin\theta_{n+1}-\sin\theta_{n})\right]^2,
 \label{eq:ell_cont2}
\end{equation}
where we have defined $\eta_n=(u_{n+1}-u_n)/b$. We also define $\phi_n =r\sin\theta_n$.
Assuming the conitnuum limit of $N\rightarrow \infty$ and $b\rightarrow 0$ while keeping the system size $L=bN$ finite, we consider 
the field variables $\theta_n \rightarrow \theta(x), u_n \rightarrow u(x)$ and  $\eta_n \rightarrow \eta(x) = \partial_x u = u^{\prime}$, where $nb\rightarrow x$ and $b \rightarrow dx$. 
We then explore the effective elasitc energy $E_{\rm el}$ as a functional of $\eta(x)$ and $\phi(x)$. 
Approximating $\theta_n \rightarrow \theta(x)-\frac{b}{2}\partial_x\theta(x)$ and $\theta_{n+1} \rightarrow \theta(x)+\frac{b}{2} \partial_x\theta(x)$ and so on, 
we may find $\cos(\theta_{n+1}+\theta_n) \rightarrow \cos(2\theta)$ and $\cos\theta_{n+1}+\cos\theta_n \rightarrow 2\cos\theta$.
Assuming that the spacial graidents of $\phi$ and $u$ are sufficiently small so that $(u^{\prime})^2$ and $(\phi^{\prime})^2$ are neglected compared to 
the 1st order contributions of $u^{\prime}(=\eta)$ and $\phi$, we obtain
\begin{equation}
 L_{A}^2-a^2  \simeq  2b^2\eta- 2b^2 (1+\eta)\phi^{\prime}+16 (\phi_0^2-\phi^2),
 \quad
 L_B^2 -b^2 \simeq 2b^2\eta+2b^2(1+\eta)\phi^{\prime}.
\end{equation}
Thus, in the continuum limit, the spring extensions can be written as 
\begin{equation}
 e_A(x) \simeq b\left[\frac{b}{a} \eta-\frac{b}{a}(1+\eta)\phi^{\prime}+\frac{8}{ab}(\phi_0^2-\phi^2)\right]
 \quad
 \mbox{and}
 \quad
 e_B(x) \simeq b\left[\eta+(1+\eta)\phi^{\prime}\right].
 \label{eq:e_a_e_b}
\end{equation}
The total elastic energy, Eq.~(\ref{eq:E-spring}),  can be written as
\begin{eqnarray}
 E_{\rm el}[\eta,\phi] &=& \int_0^L \mathcal{E} [\eta,\phi] dx,
 \quad
 \mbox{where}
 \quad
 \mathcal{E} [\eta,\phi] = \frac{k}{2b}\left(e_A^2+e_B^2\right),
\end{eqnarray}
and $L=bN$ (defined above) denotes the end-to-end distance of the ladder at its ground-state configuration.
Substituting Eqs.~(\ref{eq:e_a_e_b}) into this, we obtain
\begin{eqnarray}
 \mathcal{E} [\eta,\phi] &=& \frac{\mu}{2} \left\{\left[\frac{b}{a} \eta-\frac{b}{a}(1+\eta)\phi^{\prime}+V(\phi)\right]^2+\left[\eta+(1+\eta)\phi^{\prime}\right]^2\right\},
 \label{eq:E-density2}
\end{eqnarray}
where $\mu=kb$ has the stiffness parameter that has the dimension of (force), i.e., (energy)/(length), and we have defined 
\begin{eqnarray}
 V(\phi) &=& \frac{8}{ab}(\phi_0^2-\phi^2).
 \label{eq:V}
\end{eqnarray}
Note that the energy density is invariant under the rigid body translation, $u(x)\rightarrow u_0+u(x)$, where $u_0$ is an arbitrary constant.} \\

\textcolor{black}{
While the inertial dynamics may be fully described by the Lagrangian for the translational ($\eta$) and rotational ($\phi$) variables, here, we will rather proceed with our discussion by making assumptions that simplify the following analytical calculations. We will consider the two types of such simplifying assumption separately. \\
}

\textcolor{black}{
In the first case, we assume that $u$ (or $\eta=u^{\prime}$) goes to its equilibrium quickly enough for given $\phi$.
Solving $\partial \mathcal{E}/\partial \eta=0$, we find
\begin{eqnarray}
 \eta^{\ast}(\phi) &=& -\frac{\phi^{\prime}(1+\phi^{\prime})-\frac{b}{a}(1-\phi)\left(\frac{b}{a}\phi^{\prime}-V(\phi)\right)}{(1+\phi^{\prime})^2+\frac{b^2}{a^2}(1-\phi^{\prime})^2}.
 \label{eq:eta-ast}
\end{eqnarray}
Substituting this back into Eq.~(\ref{eq:E-density2}), and again neglecting higher order terms of $\phi^{\prime}$, 
we obtain
\begin{eqnarray}
 \mathcal{E}_{\rm eff}[\phi(x)] &\simeq& \frac{\tilde{\mu}}{2}\left[\left(V(\phi)-\frac{2b}{a}\phi^{\prime}\right)^2+\frac{4b^2}{a^2+b^2}V^2(\phi)\phi^{\prime}\right],
 \label{eq:E-density-final}
\end{eqnarray}
where we have defined the effective stiffness parameter given by $\tilde{\mu}=a^2\mu/(a^2+b^2)$ and we have neglected the terms, $V(\phi)^2{\phi^{\prime}}^2$ and $V(\phi){\phi^{\prime}}^2$.
The effective energy $E_{\rm eff}$ in Eq.~(\ref{eq:E-density-final}) is essentially that of the nonlinear Klein-Gordon (or $\phi^4$) model, with
the surface term proportional to $\phi^{\prime}$. 
Such a surface term does not change the bulk dynamics (i.e., the equations of motion), but, as demonstrated in Ref.~\cite{Chen-PNAS-2014}, is still important for the present study
as it is related to the topological charge of the soliton that energetically discriminates the anitikink from the kink. 
}

\textcolor{black}{
Here we look at the staic kink solution. The viscous dynamics will be discussed in SM Sec.~\ref{sec:viscous}.
The Euler-Lagrange equation gives
\begin{eqnarray}
 \frac{4b^2}{a^2}\phi^{\prime\prime}-V(\phi)\frac{dV(\phi)}{d\phi} &=& 0.
 \label{eq:EL_eq}
\end{eqnarray}
Under the boudary conditions given by $\phi(|x|\rightarrow\infty)=\pm \phi_0$ and $\phi^{\prime}(|x|\rightarrow \infty) =0$, Eq.~(\ref{eq:EL_eq}) can be integrated once to obtain
\begin{eqnarray}
 \frac{2b}{a}\phi^{\prime} &=& \pm V(\phi)
\end{eqnarray}
The positive one, $(2b/a)\phi^{\prime}=V(\phi)$, gives a kink solution, $\phi_{\rm kink}$, that gives the square braket term in Eq.~(\ref{eq:E-density-final}) precisely zero, 
while the negative one, $(2b/a)\phi^{\prime}=-V(\phi)$, gives an "antikink" solution, $\phi_{\rm ghost}$, which has a higher energy.
These solutions are explicitly given by
\begin{equation}
 \phi_{\rm kink} \ = \ \phi_0\tanh (x/\xi),
 \qquad
 \phi_{\rm ghost} \ = \ -\phi_0\tanh(x/\xi)
\end{equation}
where
\begin{equation}
 \phi_0 \ = \ \frac{s}{2}\sin\alpha, 
 \qquad
 \xi  \ = \ \frac{b^2}{2 s \sin\alpha}
\end{equation}
It is important to note that the higher-energy solution, $\phi_{\rm ghost}$, is a "ghost" distinct from an antikink that we actually observe in the Jacob's ladder, as this ghost assumes the straight configuration of the green linkages, which never happens.
The physically observable antikink will be briefly discussed in the next subsection.} \\

\textcolor{black}{
In the second approach, we entirely neglect $\eta$ and simply set $\eta(x)=0$ in $E_{\rm eff}$, which leads to
\begin{eqnarray}
 \mathcal{E}_{\rm eff}[\phi(x)] &=& \frac{\mu}{2}\left[\left(V(\phi)-\frac{b}{a}\phi^{\prime}\right)^2+{\phi^{\prime}}^2\right].
 \label{eq:E-density-reduced}
\end{eqnarray}
The kink solution is again given by $\phi_{\rm kink}(x)= \phi_0 \tanh(x/\xi_{0})$, where the interface width $\xi_0$ is slightly changed as 
\begin{eqnarray}
 \xi_0 &=& \frac{b\sqrt{a^2+b^2}}{4s\sin\alpha}, 
 \quad
 \mbox{or}
 \quad
 \xi = \frac{2b}{\sqrt{a^2+b^2}}\xi_0.
\end{eqnarray}
Thus, the coupling with $\eta$ changes the spatial profile of $\phi$ slightly as well as the surface energy terms in $E_{\rm eff}$, but 
the physics is essentially unchaged from the simpler system considered here. 
Finally, for a symmetric Jacob's ladder, $a/b\rightarrow 1$ (and $\alpha \rightarrow 90^{\circ}$), we find 
\begin{eqnarray}
 \phi_{\rm kink} &=&  +\frac{s}{2} \tanh\left[\frac{2\sqrt{2}s}{b^2} x \right].
 \label{eq:kink-reduced}
\end{eqnarray}
}

\textcolor{black}{
\subsection{Antikink propagation}
Due to essentially floppy nature of the ladder, the green links can also rotate (in addtion to the red ones), which allows us to observe the antikink propagation in the 
same direction as the kink for large $\alpha$. See Supplemental Movie 5 and Fig.~3 (b) in the main text.
For small $\alpha$, however, we have seen that antikink behavior is more involved. See Fig.~\ref{fig:antikink-50} in SM Sec.~\ref{sec:simulation}.
It is beyond the scope of our present study to develop a full continuum theory that can precisely describe the antikink propagation observed in numerical simulations, and we leave this subject to our future studies.
}

\textcolor{black}{ 
\subsection{Viscous dynamics}~\label{sec:viscous}
Our interest primarly lies in the viscous dynamics where the gravitational driving and frictional forces from a surroudning medium (water) balace and the nonlinear wave
propagates along the ladder at an apparently constant speed, $c_s$, as observed in the water tank experiment.
Note that this is diferent from the inerital regime stuided previously for which the solitary wave shows the Lorentz contraction.
}

\textcolor{black}{
As we have seen above, the translational degree of freedom $\eta$ modifies the interface width but does not change the physics qualitatively.
From this, here we will simply assume $\eta(x)=0$ from the begging.
Then  the local velocity of node A relative to B is ${\boldsymbol v}_{A, n}=s\dot{\theta}_n (\hat{\boldsymbol e}_z \times \hat{\boldsymbol t}_n)$, from which the kinetic energy of the ladder is given by
\begin{eqnarray}
 E_{\rm kin} &=& \sum_{n=1}^N \frac{1}{2}m s^2 \dot{\theta}_n^2  \rightarrow \int_0^L \frac{I}{2} \frac{\dot{\phi}^2}{r^2-\phi^2} dx,
  \label{eq:K}
 \end{eqnarray}
where $I=ms^2/b$ is the moment of inertia (per unit length) of a block about the node B. 
}

\textcolor{black}{    
Within the assumption made here, the gravitational potential energy should be considered only for the A nodes:
\begin{eqnarray}
 U_g &=& -\sum_{n=1}^N mg x_{A,n} \simeq  -\sum_{n=1}^N mg (x_{B,n}-s\sin\theta_n) \simeq -\frac{mgL^2}{2b}\left(1+\frac{r}{L}\sin\alpha\right)+\int_0^L \frac{2mg}{b} \phi(x) dx.
 \label{eq:U_g}
\end{eqnarray}
}
\textcolor{black}{
The Lagarangian of this system is thus given by $L=E_{\rm kin}+E_{\rm el}+U_g$, i.e., 
\begin{eqnarray}
 L &=& \int_0^L  dx\left[\frac{I}{2} \frac{\dot{\phi}^2}{r^2-\phi^2}-\frac{\mu}{2}\left\{\left(V(\phi)-\frac{b}{a}\phi^{\prime}\right)^2+{\phi^{\prime}}^2\right\}-\frac{2mg}{b}\phi \right]
\end{eqnarray}
}

\textcolor{black}{
Assuming the drag force that is linearly proportional to the local velocity of node A, we consider the Rayleigh dissipation function given by
\begin{eqnarray}
 R &=& \sum_{n=1}^N \frac{\zeta}{2} v^2_{A,n} \rightarrow \int_0^L \frac{\zeta s^2}{2b}\frac{\dot{\phi}^2}{r^2-\phi^2} dx.
 \label{eq:Rayleigh}
\end{eqnarray}
}

\textcolor{black}{
The time evolution of $\phi$ can be described by  the Euler-Lagrangian equation supplemented with the non-conservative friction force as
\begin{eqnarray}
 \frac{d}{dt}\left(\frac{\partial L}{\partial \dot{\phi}}\right)+\frac{d}{dx}\left(\frac{\partial L}{\partial {\phi}^{\prime}}\right)-\frac{\partial L}{\partial \phi} &=& -\frac{\partial R}{\partial \dot{\phi}}.
\end{eqnarray}
Assuming $\phi^2\ll r^2$,  we may obtain
\begin{eqnarray}
 \frac{I}{r^2}\ddot{\phi} &=& -\frac{\zeta s^2}{br^2}\dot{\phi}+\mu \left(1+\frac{b^2}{a^2}\right)\phi^{\prime\prime}-\mu V(\phi)\frac{dV(\phi)}{d\phi}-\frac{2mg}{b}.
 \label{eq:eqm-kink}
\end{eqnarray}
Rescaling all the relevant variables according to 
\begin{equation}
 T = t/(2\tau), 
 \qquad
 X = x/b,
 \qquad
 f = \phi/\phi_0,
\end{equation}
where the unit of time is chosen as $\tau = \sqrt{m/k}$, we have
\begin{eqnarray}
 \frac{\partial^2f}{\partial T^2} &=& -\Gamma \frac{\partial f}{\partial T} +K\frac{\partial^2 f}{\partial X^2}+A f(1-f^2)-\epsilon,
 \label{eq:eqm-kink-rescaled}
\end{eqnarray}
where 
\begin{equation}
 \Gamma = \frac{2\zeta \tau}{m},
 \qquad
 K = 1+\frac{b^2}{a^2}, 
 \qquad
 A = \frac{32 s^2}{a^2}\sin^2\alpha,
 \qquad
 \epsilon = \frac{4g\tau^2}{s \sin\alpha} = \frac{4mg}{ks \sin\alpha}.
\end{equation}
Note that the non-dimensional external force $\epsilon$ is very small because $mg \ll ks$.
}

\textcolor{black}{
Since we are interested in the "overdamped" regime given by $T \gg \Gamma^{-1}$, we neglect the inertia term in Eq.~(\ref{eq:eqm-kink-rescaled}), and 
investigate the following equation
\begin{eqnarray}
 \Gamma \frac{\partial f}{\partial T}  &=& K\frac{\partial^2 f}{\partial X^2}+A f(1-f^2)-\epsilon.
 \label{eq:eqm-kink-rescaled-2}
\end{eqnarray}
For $0<\epsilon \ll 1$, Eq.~(\ref{eq:eqm-kink-rescaled-2}) has an approximate analytic solution valid to the 1st order of $\epsilon$ given by
\begin{eqnarray}
 f(X,T) &=& -\frac{\epsilon}{2A}+\tanh \left(\sqrt{\frac{A}{2K}} X-\frac{3\epsilon}{2\Gamma} T\right),
 \label{eq:kink-sol}
\end{eqnarray}
which represents a propagating kink at a constant speed towards the positive $x$ direction (the direction of the gravity), as observed in the experiment and in the numerical simulation.
The phase of Eq.~(\ref{eq:kink-sol}), $\Theta$, can be written in the dimensionful form as
\begin{eqnarray}
 \Theta(x,t) &=& \sqrt{\frac{A}{2K}} X-\frac{3\epsilon}{2\Gamma} T = \frac{4s\sin\alpha}{b\sqrt{a^2+b^2}}x- \frac{3mg}{2\zeta s \sin\alpha} t.
\end{eqnarray}
The phase speed, $c_{\rm kink}$, is thus found from $d\Theta=0$ to be [Eq.~(4) in the main text]
\begin{eqnarray}
 c_{\rm kink}(\alpha) &=& \frac{dx}{dt} = \frac{3b\sqrt{a^2+b^2}}{8 s^2 \sin^2\alpha}\frac{mg}{\zeta}.
 \label{eq:c_s}
\end{eqnarray}
While the scaling, $c_{\rm kink}\sim mg/\zeta$, is intuitively easy to understand from the force balance, the geometric relation, $c_{\rm kink}\sim (\sin\alpha)^{-2}$, seems non-trivial. 
}

\textcolor{black}{
To check this analytical prediction with numerical data, we look at the kink position $X_{\rm kink}$ as  a function of time $t$ in our numerical simulation.
For this purpose, we propose that the kink position $X_{\rm kink}$ can be calculated via 
\begin{eqnarray}
 X_{\rm kink} &=& \frac{\displaystyle{\sum_{n=0}^{N-1} \frac{1}{2}(X_{n+1}+X_n)(\phi_{n+1}-\phi_n)^2}}{\displaystyle{\sum_{n=0}^{N-1}  (\phi_{n+1}-\phi_n)^2}},
 \label{eq:X_d}
\end{eqnarray}
which turns out to work surprisingly well when the system contains a single kink only. See Fig.~3 (c) and (d) in the main text.
}

\section{Scaled experiment}
\begin{figure}[htbp]
\begin{center}
 \includegraphics[width=0.70\linewidth]{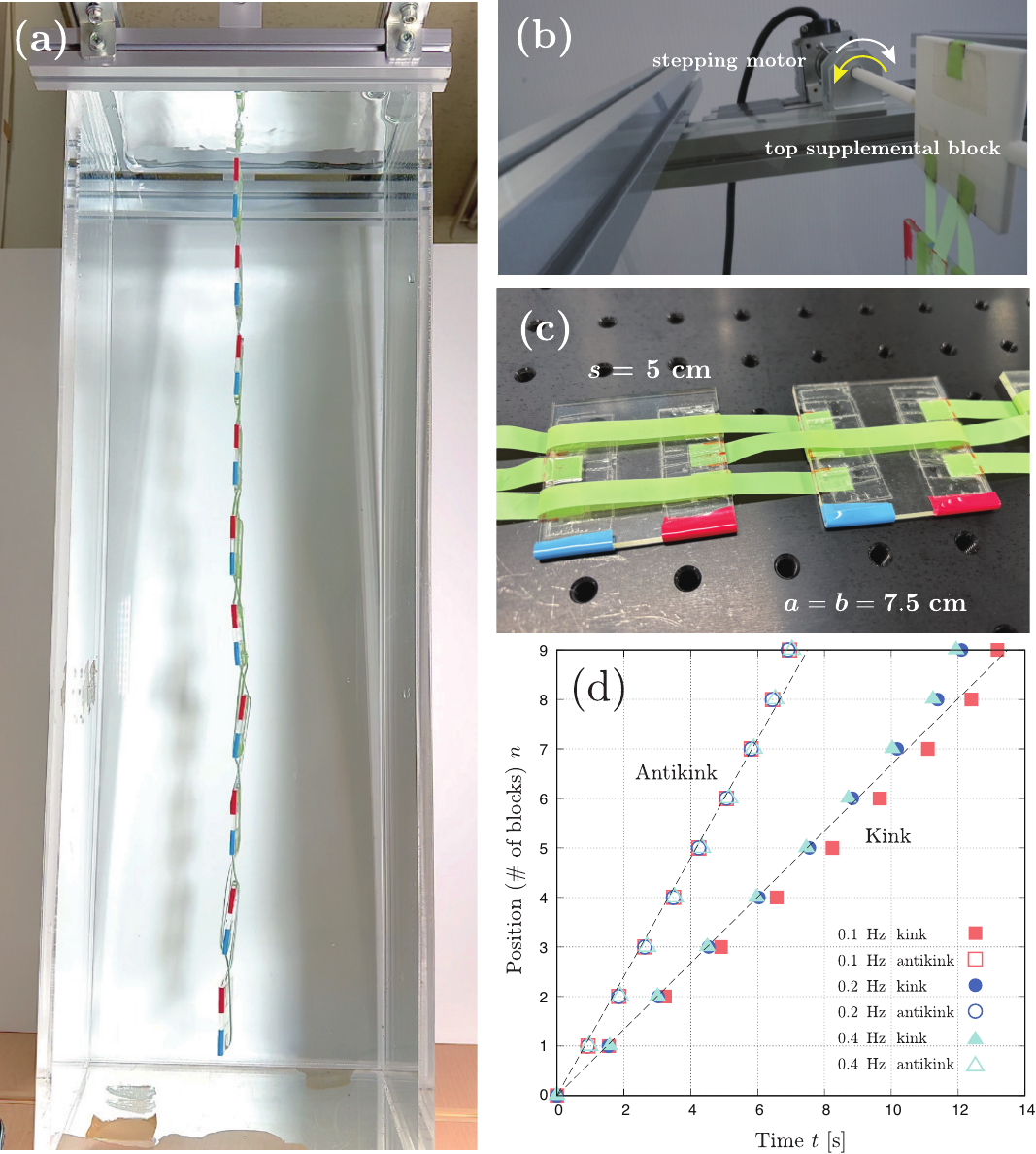}
\caption{Schematics of our experimental apparatus: photographs and their demensions. (a) A water tank (b) An active rotator system based on the stepping motor (c) Manufactured plastic Jacob's ladder immersed in the water for experimental observations. (d) Kink and antikink speeds measured in the water tank experiment for the "2-1" connection $N=10$ ladder, for different angular velocities of the stepping motor, 0.1, 0.2, 0.4 Hz.}
\label{fig:sm_experiment}
\end{center}
\end{figure}
Experiments were done in an acrylic tank with dimensions 270 mm $\times$ 270 mm and a water level of 800 mm filled with tap water, as shown in Fig.~\ref{fig:sm_experiment} (a).
Figure~\ref{fig:sm_experiment} (b) is the photograph of the hand-made plastic Jacob's ladder consisting of $N=10$ acrylic blocks of size $s \times s$, where  $s=5$ cm, and thickness $t=3$ mm, which was immersed in the water. 
The blocks are conncted by the tapes of OPP film of thickness $40\, \mu$m, so that the neighboring blocks have the approximately uniform distance of $b=7.5$ cm.
Considering the relative density of the acrylic block is $\rho/\rho_{\rm w}=1.2$, where $\rho_{\rm w}\approx 10^3$ kg/m$^3$ is the density of water, the effective gravitional acceleration in water $g^{\ast}=(1-\rho_{\rm w}/\rho)g$ is just about 10\% of $g$. 
We used the stepping motor to rotate the top supplemental block, which itself is locating above the water surface; See Fig.~\ref{fig:sm_experiment} (c). Note that this supplemental block is not included 
in the main part of the ladder consisting of $N=10$ blocks. 
That is, our ladder consists of the total 11 blocks, with $N=10$ blocks immersed in the water. 

In Fig.~\ref{fig:sm_experiment} (d), we plot the experimentally observed kink (and anitkink) position as a function of time $t$ for the increasing rotational excitation speeds of the initial block above the free water surface [Fig.~\ref{fig:sm_experiment} (c)]. 
When the $n$-th block tilts at angle $90^{\circ}$ (i.e., when it gets horizontal), we record $t_n$ as the time at which the kink (and antikink) reaches the $n$-th block.
We set $t_0=0$.
The data points shown represent the averages over five independent measurements, with their error bars smaller than the symbols.
For all the cases investigated here, though there appears to be a slight tendency towards acceleration in the larger $n$ region, we overall see that the kink (and antikink) moves with an approximately constant speed, which we denote $c_{\rm kink}$ (and $c_{\rm antikink})$.
The wave speeds are insensitive to the initial angular velocity, because the friction from the surrounding media and the gravitational driving balance to realize the constant 
speed kink propagation in our system. See SM Sec.~\ref{sec:viscous}.
Thereotically, for the symmetric design $a=b$ studied in this experiment, $c_{\rm kink}=c_{\rm antikink}$ should hold, which has been confirmed
in the numerical simulation for $\alpha=87^{\circ}$ [Supplemental Movies 6 and 7].
The difference in the wave speeds is thus observd only in our experiment, which is further discussed in SM Sec.~\ref{sec:asymmetry}.
\begin{figure}[htbp]
\begin{center}
 \includegraphics[width=0.20\linewidth]{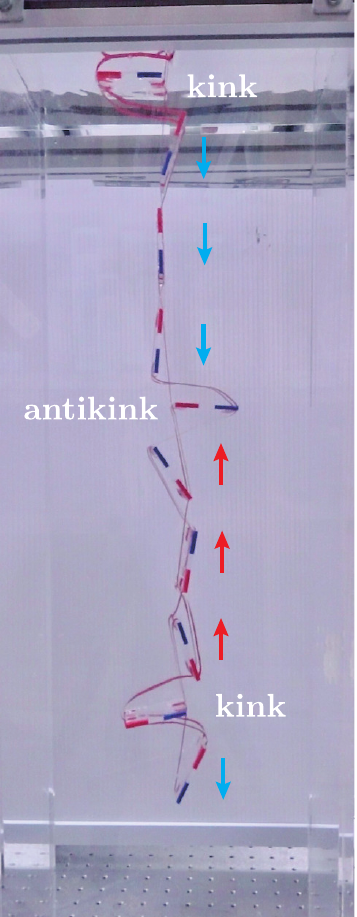}
\caption{Multiple kinks unidirectionally propagating down the ladder observed in the water-tank experiment. Arrows indicate the orientations of the blocks. There are three domain walls.}
\label{fig:multi-kinks}
\end{center}
\end{figure}
\\

\textcolor{black}{
In principle, if this floppy chain is inifinitely long, it could support arbitrary number of kinks and antikinks.
To demonstrate this, we tried creating as many as domain walls in our chain in the water-tank experiment. 
At best, we were able to observe a maximum of three domain walls, i.e., kink, antikink, and kink, as shown in Figure~\ref{fig:multi-kinks}.
Because of the practical length limitation, we were not successful to creating more than four waves. 
}

\textcolor{black}{
\section{Wave speed asymmetry}~\label{sec:asymmetry}
In the water-tank experiment using $\alpha \simeq 90^{\circ}$ ladder, we created a kink and antikink separately to find $c_{\rm antikink}/c_{\rm kink} = 1.75$. 
See Fig.~\ref{fig:sm_experiment} (c).
We thus experimentally confirm that $c_{\rm antikink}/c_{\rm kink} > 1$ is not due to the kink-antikink interaction. 
}

\textcolor{black}{
Most probably, the origin of this unexpected asymmetry in the wave speed is attributable to the limitation of our fabrication precision.
We have found that the preferential curvatures in the OPP films used at hinges slightly deform
the bistable energy landscape, with one stable configuration having a lower energy than the other.
}

\begin{figure}[htbp]
\begin{center}
 \includegraphics[width=0.95\linewidth]{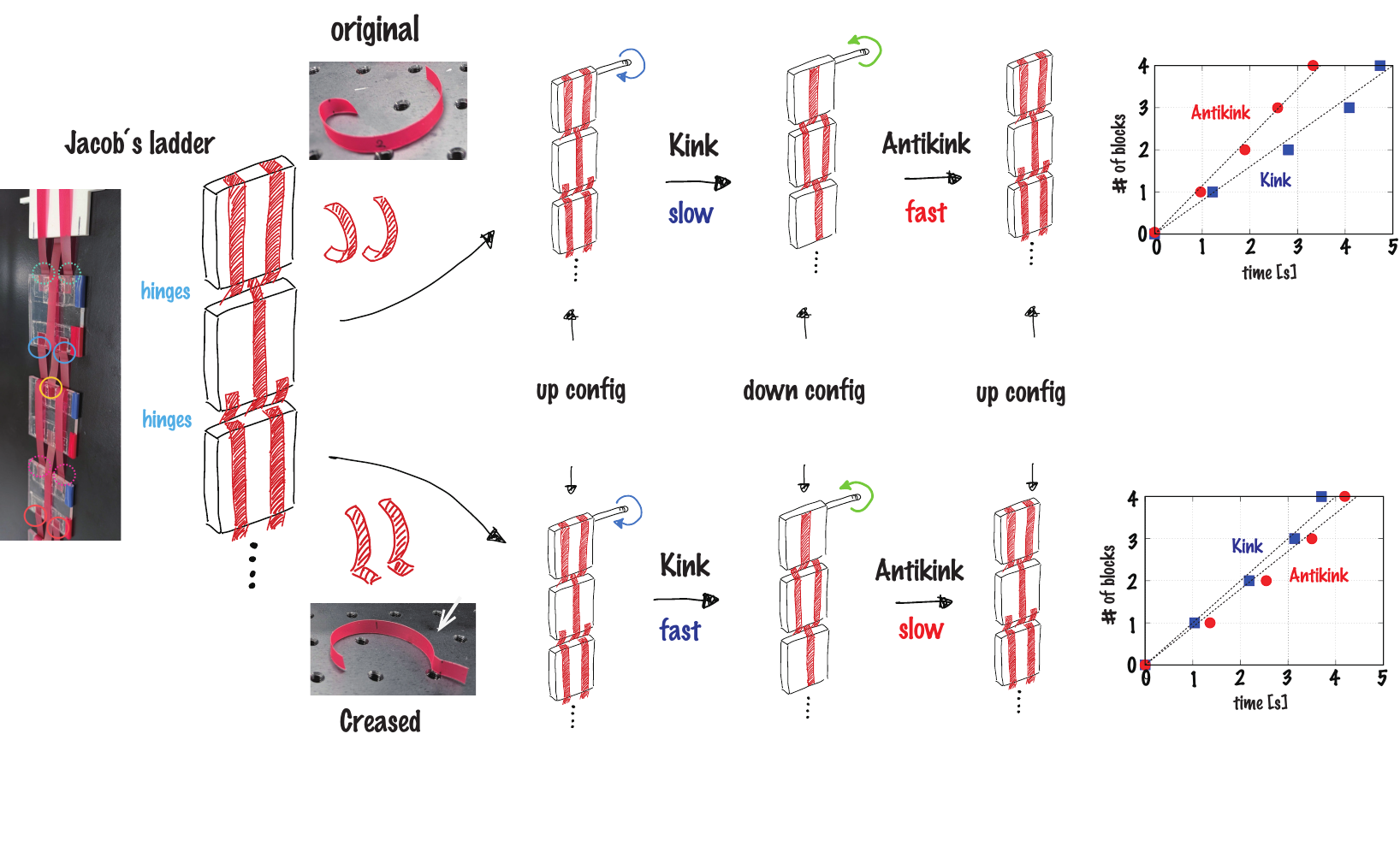}
\caption{Experimental procedure to test the effects of spontaneous curvatures of the tapes at the hinges on the wave speeds. The ladder used here consists of $N=5$ blocks.
Each data point is the average over three independent measurements with the error bar smaller than the symbol size.}
\label{fig:protocol}
\end{center}
\end{figure}

\textcolor{black}{
To test this hypothesis, we intentionally creased the portion of the tapes corresponding to the hinges, thereby introducing a tendency for the tapes to bend in the opposite direction of its original curvature. 
We then conducted the same experiments and find that the relationship between the propagation speeds of kinks and anti-kinks was indeed reversed. See Fig.~\ref{fig:protocol}.
This suggests that, because the hinge section bends more easily in one direction than in the opposite direction, each block rotates more easily in one direction than in the other direction.
This may cause a difference in the speeds of the two waves, which should otherwise be identical. 
As we have not yet exhaused all the possibilities, we cannot definitively conclude that the tape's intrinsic curvature alone is the origin of the asymmetry, but it is undoubtedly one of the primary causes.
}

\textcolor{black}{
Because a block has a finite thickness, the tape connecting the blocks is constantly in a bent state at the hinge. 
Therefore, when the ladder model is left undisturbed on a table for a few days, the tape tends to develop its spontaneous curvature in the direction of the bend imposed. 
Thus, the model energetically prefers the configuration it assumes when placed on the table over the other of its two bistable configurations. 
As a result, the wave traveling from the original configuration to the other moves more slowly, while the wave traveling from the other configuration back to the original one moves faster. 
We verified this observation by conducting experiments in which we compared the ratios of the wave speeds by varying the configurations of the model when placed on the table. 
The result further supports the conclusion that the spontaneous curvature of the tape at the hinge is the cause of the wave speed asymmetry.
}

\textcolor{black}{
Note that the asymmetry can be observed for sufficiently small gravitational driving $g^{\ast}/g\ll1$ such as in our water tank experiment. 
In air, the gravity is strong enough so that the block inertia dominates, washing out the subtle differnece of the two wave speeds, which is probably the reason why we never notice it usually.
}

\section*{Captions of supplemental movies}

\noindent
{\bf SI Movie 1:}
 This video shows how a commercially available wooden Jacob's ladder toy ("Pata-pata" in Japanese) works in air. Real-time video replay.\\

\noindent
{\bf SI Movie 2:}
 Wave propagation in custum made ladder of $N=10$ and $\alpha\simeq 90^{\circ}$ observed in water-tank experiment. Real-time video replay.\\

\noindent
{\bf SI Movie 3:}
 A video rotated $90$ degrees that focuses on the movement of the boards in a adder of $N=10$ and $\alpha\simeq 90^{\circ}$ observed in water-tank experiment. Real-time video replay.\\

\noindent
{\bf SI Movie 4:}
Kink wave propagation in numerical simulation for $N=10$ and $\alpha=60^{\circ}$. \\

\noindent
{\bf SI Movie 5:}
Antikink wave propagation in numerical simulation for $N=10$ and $\alpha=60^{\circ}$.\\

\noindent
{\bf SI Movie 6:}
Kink wave propagation in numerical simulation for $N=10$ and $\alpha=87^{\circ}$. \\

\noindent
{\bf SI Movie 7:}
Antikink wave propagation in numerical simulation for $N=10$ and $\alpha=87^{\circ}$. \\

\noindent
{\bf SI Movie 8:}
Kink wave propagation in numerical simulation for $N=10$ and $\alpha=50^{\circ}$. \\

\noindent
{\bf SI Movie 9:}
Antikink wave propagation in numerical simulation for $N=10$ and $\alpha=50^{\circ}$. \\

\noindent
{\bf SI Movie 10:}
Antikink stops halfway through a 3D printed rigid-linked asymmetric ladder under gravity in air. Real-time video replay.\\

\noindent
{\bf SI Movie 11:}
Excitation of kink and antikink in water-tank experiment. Real-time video replay.\\

\noindent
{\bf SI Movie 12:}
Pair annihilation of kink and antikink in water-tank experiment. Real-time video replay. \\

\noindent
{\bf SI Movie 13:}
Spin wave observed in a dual ring chain, with one strand fixed with gluing tapes so that the rings in that strand do not rotate. Video frame rate 240 fps. \\

\noindent
{\bf SI Movie 14:}
Head-on collision between kink and antikink within a suspended chain in air. A 3D-printed rigid-link Jacob's ladder of $\alpha\simeq 90^{\circ}$ consists of $N=17$ blocks. Real-time video replay.



\begin{thebibliography}{43}%
\makeatletter
\providecommand \@ifxundefined [1]{%
 \@ifx{#1\undefined}
}%
\providecommand \@ifnum [1]{%
 \ifnum #1\expandafter \@firstoftwo
 \else \expandafter \@secondoftwo
 \fi
}%
\providecommand \@ifx [1]{%
 \ifx #1\expandafter \@firstoftwo
 \else \expandafter \@secondoftwo
 \fi
}%
\providecommand \natexlab [1]{#1}%
\providecommand \enquote  [1]{``#1''}%
\providecommand \bibnamefont  [1]{#1}%
\providecommand \bibfnamefont [1]{#1}%
\providecommand \citenamefont [1]{#1}%
\providecommand \href@noop [0]{\@secondoftwo}%
\providecommand \href [0]{\begingroup \@sanitize@url \@href}%
\providecommand \@href[1]{\@@startlink{#1}\@@href}%
\providecommand \@@href[1]{\endgroup#1\@@endlink}%
\providecommand \@sanitize@url [0]{\catcode `\\12\catcode `\$12\catcode
  `\&12\catcode `\#12\catcode `\^12\catcode `\_12\catcode `\%12\relax}%
\providecommand \@@startlink[1]{}%
\providecommand \@@endlink[0]{}%
\providecommand \url  [0]{\begingroup\@sanitize@url \@url }%
\providecommand \@url [1]{\endgroup\@href {#1}{\urlprefix }}%
\providecommand \urlprefix  [0]{URL }%
\providecommand \Eprint [0]{\href }%
\providecommand \doibase [0]{https://doi.org/}%
\providecommand \selectlanguage [0]{\@gobble}%
\providecommand \bibinfo  [0]{\@secondoftwo}%
\providecommand \bibfield  [0]{\@secondoftwo}%
\providecommand \translation [1]{[#1]}%
\providecommand \BibitemOpen [0]{}%
\providecommand \bibitemStop [0]{}%
\providecommand \bibitemNoStop [0]{.\EOS\space}%
\providecommand \EOS [0]{\spacefactor3000\relax}%
\providecommand \BibitemShut  [1]{\csname bibitem#1\endcsname}%
\let\auto@bib@innerbib\@empty
\bibitem [{\citenamefont {Klein}\ and\ \citenamefont
  {Sommerfeld}(2010)}]{Klein-Book}%
  \BibitemOpen
  \bibfield  {author} {\bibinfo {author} {\bibfnamefont {F.}~\bibnamefont
  {Klein}}\ and\ \bibinfo {author} {\bibfnamefont {A.}~\bibnamefont
  {Sommerfeld}},\ }\href@noop {} {\emph {\bibinfo {title} {{The Theory of the
  Top. Volume II: Development of the Theory in the Case of the Heavy Symmetric
  Top}}}}\ (\bibinfo  {publisher} {{Birkh\"{a}user}},\ \bibinfo {address}
  {Boston, MA},\ \bibinfo {year} {2010})\BibitemShut {NoStop}%
\bibitem [{\citenamefont {Bondi}(1986)}]{Bondi-PRLA-1986}%
  \BibitemOpen
  \bibfield  {author} {\bibinfo {author} {\bibfnamefont {H.}~\bibnamefont
  {Bondi}},\ }\bibfield  {title} {\bibinfo {title} {{The rigid body dynamics of
  unidirectional spin}},\ }\href@noop {} {\bibfield  {journal} {\bibinfo
  {journal} {Proc. R. Soc. Lond. A}\ }\textbf {\bibinfo {volume} {405}},\
  \bibinfo {pages} {265} (\bibinfo {year} {1986})}\BibitemShut {NoStop}%
\bibitem [{\citenamefont {Kondo}\ and\ \citenamefont
  {Nakanishi}(2017)}]{Kondo-PRE-2017}%
  \BibitemOpen
  \bibfield  {author} {\bibinfo {author} {\bibfnamefont {Y.}~\bibnamefont
  {Kondo}}\ and\ \bibinfo {author} {\bibfnamefont {H.}~\bibnamefont
  {Nakanishi}},\ }\bibfield  {title} {\bibinfo {title} {{Rattleback dynamics
  and its reversal time of rotation}},\ }\href@noop {} {\bibfield  {journal}
  {\bibinfo  {journal} {Phys. Rev. E}\ }\textbf {\bibinfo {volume} {95}},\
  \bibinfo {pages} {062207} (\bibinfo {year} {2017})}\BibitemShut {NoStop}%
\bibitem [{\citenamefont {Moffatt}(2000)}]{Moffatt-Nature-2000}%
  \BibitemOpen
  \bibfield  {author} {\bibinfo {author} {\bibfnamefont {H.~K.}\ \bibnamefont
  {Moffatt}},\ }\bibfield  {title} {\bibinfo {title} {{Euler's disk and its
  finite-time singularity}},\ }\href@noop {} {\bibfield  {journal} {\bibinfo
  {journal} {Nature}\ }\textbf {\bibinfo {volume} {404}},\ \bibinfo {pages}
  {833} (\bibinfo {year} {2000})}\BibitemShut {NoStop}%
\bibitem [{\citenamefont {Sekimoto}\ \emph {et~al.}(2021)\citenamefont
  {Sekimoto}, \citenamefont {Benane}, \citenamefont {Alloubia}, \citenamefont
  {Arteil},\ and\ \citenamefont {Fruleux}}]{Sekimoto-EPJB-2021}%
  \BibitemOpen
  \bibfield  {author} {\bibinfo {author} {\bibfnamefont {K.}~\bibnamefont
  {Sekimoto}}, \bibinfo {author} {\bibfnamefont {Y.~M.}\ \bibnamefont
  {Benane}}, \bibinfo {author} {\bibfnamefont {K.~E.}\ \bibnamefont
  {Alloubia}}, \bibinfo {author} {\bibfnamefont {R.}~\bibnamefont {Arteil}},\
  and\ \bibinfo {author} {\bibfnamefont {A.}~\bibnamefont {Fruleux}},\
  }\bibfield  {title} {\bibinfo {title} {{A mechanism of macroscopic rigid-body
  behavior through evanescent mode}},\ }\href@noop {} {\bibfield  {journal}
  {\bibinfo  {journal} {Euro. Phys. J. B}\ }\textbf {\bibinfo {volume} {94}},\
  \bibinfo {pages} {106} (\bibinfo {year} {2021})}\BibitemShut {NoStop}%
\bibitem [{\citenamefont {\textcolor{red}{Tadashi
  Tokieda}}(2026)}]{Tokieda-Book-2026}%
  \BibitemOpen
  \bibfield  {author} {\bibinfo {author} {\bibnamefont {\textcolor{red}{Tadashi
  Tokieda}}},\ }\href@noop {} {\emph {\bibinfo {title} {\textcolor{red}{A
  Child's Eye, an Adult's Mind: Playing with Mathematics}}}}\ (\bibinfo
  {publisher} {\textcolor{red}{Nippon Hyoron Sha}},\ \bibinfo {address}
  {\textcolor{red}{Tokyo}},\ \bibinfo {year} {\textcolor{red}{2026}})\ \bibinfo
  {note} {\textcolor{red}{[in Japanese]}}\BibitemShut {NoStop}%
\bibitem [{\citenamefont {\textcolor{red}{M. Saad Bhamla and B. Benson and C.
  Chai and G. Katsikis and A. Johri and M.
  Prakash}}(2017)}]{Bhamla-NatBiomed-2017}%
  \BibitemOpen
  \bibfield  {author} {\bibinfo {author} {\bibnamefont {\textcolor{red}{M. Saad
  Bhamla and B. Benson and C. Chai and G. Katsikis and A. Johri and M.
  Prakash}}},\ }\bibfield  {title} {\bibinfo {title}
  {\textcolor{red}{Hand-powered ultralow-cost paper centrifuge}},\ }\href@noop
  {} {\bibfield  {journal} {\bibinfo  {journal} {\textcolor{red}{Nat. Biomed.
  Eng.}}\ }\textbf {\bibinfo {volume} {\textcolor{red}{1}}},\ \bibinfo {pages}
  {\textcolor{red}{0009}} (\bibinfo {year}
  {\textcolor{red}{2017}})}\BibitemShut {NoStop}%
\bibitem [{\citenamefont {Edge}(1998)}]{Edge-PhysTeach-1998}%
  \BibitemOpen
  \bibfield  {author} {\bibinfo {author} {\bibfnamefont {R.}~\bibnamefont
  {Edge}},\ }\bibfield  {title} {\bibinfo {title} {{Solitons}},\ }\href@noop {}
  {\bibfield  {journal} {\bibinfo  {journal} {Phys. Teach.}\ }\textbf {\bibinfo
  {volume} {36}},\ \bibinfo {pages} {483} (\bibinfo {year} {1998})}\BibitemShut
  {NoStop}%
\bibitem [{\citenamefont {Dickens}(1850)}]{Dickens}%
  \BibitemOpen
  \bibfield  {author} {\bibinfo {author} {\bibfnamefont {C.}~\bibnamefont
  {Dickens}},\ }\bibfield  {title} {\bibinfo {title} {{A Christmas Tree}},\
  }in\ \href@noop {} {\emph {\bibinfo {booktitle} {{Household Words}}}},\
  Vol.~\bibinfo {volume} {II}\ (\bibinfo  {publisher} {{Bradbury \& Evans}},\
  \bibinfo {year} {1850})\BibitemShut {NoStop}%
\bibitem [{\citenamefont {Immel}(2026)}]{cotsen-blog}%
  \BibitemOpen
  \bibfield  {author} {\bibinfo {author} {\bibfnamefont {A.}~\bibnamefont
  {Immel}},\ }\bibfield  {title} {\bibinfo {title} {{The Jacob's Ladder Toy and
  Its Mysterious History}}} (\bibinfo {year} {2026}),\ \bibinfo {note} {{\it
  Cotsen Children's Library (Princeton University)}, Available from
  {https://cotsen.blogs.princeton.edu/2019/01/the-jacobs-ladder-toy-and-its-mysterious-history/},
  [Accessed March 21, 2026]}\BibitemShut {NoStop}%
\bibitem [{Sci(1889)}]{SciAm-1889}%
  \BibitemOpen
  \bibfield  {title} {\bibinfo {title} {{Jacob's Ladder}},\ }\href@noop {}
  {\bibfield  {journal} {\bibinfo  {journal} {Sci. Am.}\ }\textbf {\bibinfo
  {volume} {61}},\ \bibinfo {pages} {227} (\bibinfo {year} {1889})}\BibitemShut
  {NoStop}%
\bibitem [{\citenamefont {Ohta}\ and\ \citenamefont {Kitao}(1931)}]{Book-Edo}%
  \BibitemOpen
  \bibfield  {author} {\bibinfo {author} {\bibfnamefont {N.}~\bibnamefont
  {Ohta}}\ and\ \bibinfo {author} {\bibfnamefont {S.}~\bibnamefont {Kitao}},\
  }\href@noop {} {\emph {\bibinfo {title} {{Edo Nishiki}}}}\ (\bibinfo
  {publisher} {Yoneyama Doh},\ \bibinfo {address} {Tokyo},\ \bibinfo {year}
  {1931})\BibitemShut {NoStop}%
\bibitem [{SM()}]{SM}%
  \BibitemOpen
  \href@noop {} {}\bibinfo {note} {See Supplemental Material at ***** for
  further details of the experiments and extended theoretical analysis, which
  includes Ref.}\BibitemShut {Stop}%
\bibitem [{\citenamefont {Maxwell}(1864)}]{Maxwell-PhilosMag-1864}%
  \BibitemOpen
  \bibfield  {author} {\bibinfo {author} {\bibfnamefont {J.~C.}\ \bibnamefont
  {Maxwell}},\ }\bibfield  {title} {\bibinfo {title} {{On the calculation of
  the equilibrium and stiffness of frames}},\ }\href@noop {} {\bibfield
  {journal} {\bibinfo  {journal} {Philos. Mag.}\ }\textbf {\bibinfo {volume}
  {27}},\ \bibinfo {pages} {294} (\bibinfo {year} {1864})}\BibitemShut
  {NoStop}%
\bibitem [{\citenamefont {Calladine}(1978)}]{Calladine-IJSS-1978}%
  \BibitemOpen
  \bibfield  {author} {\bibinfo {author} {\bibfnamefont {C.~R.}\ \bibnamefont
  {Calladine}},\ }\bibfield  {title} {\bibinfo {title} {{Buckminster Fuller's
  "Tensegrity" structures and Clerk Maxwell's rules for the construction of
  stiff frames}},\ }\href@noop {} {\bibfield  {journal} {\bibinfo  {journal}
  {Int. J. Solid Struct.}\ }\textbf {\bibinfo {volume} {14}},\ \bibinfo {pages}
  {161} (\bibinfo {year} {1978})}\BibitemShut {NoStop}%
\bibitem [{\citenamefont {Stronge}(1987)}]{Stronge-PRSLA-1987}%
  \BibitemOpen
  \bibfield  {author} {\bibinfo {author} {\bibfnamefont {W.~J.}\ \bibnamefont
  {Stronge}},\ }\bibfield  {title} {\bibinfo {title} {{The domino effect: a
  wave of destabilizing collisions in a periodic array}},\ }\href@noop {}
  {\bibfield  {journal} {\bibinfo  {journal} {Proc. R. Soc. Lond. A}\ }\textbf
  {\bibinfo {volume} {409}},\ \bibinfo {pages} {199} (\bibinfo {year}
  {1987})}\BibitemShut {NoStop}%
\bibitem [{foo()}]{footnote-1}%
  \BibitemOpen
  \href@noop {} {}\bibinfo {note} {The Japanese name of Jacob's ladder, {\it
  Kata kata} or {\it Pata pata}, was named after its unique clattering
  sound.}\BibitemShut {Stop}%
\bibitem [{\citenamefont {Kane}\ and\ \citenamefont
  {Lubensky}(2014)}]{Kane-NatPhys-2014}%
  \BibitemOpen
  \bibfield  {author} {\bibinfo {author} {\bibfnamefont {C.~L.}\ \bibnamefont
  {Kane}}\ and\ \bibinfo {author} {\bibfnamefont {T.~C.}\ \bibnamefont
  {Lubensky}},\ }\bibfield  {title} {\bibinfo {title} {{Topological boundary
  modes in isostatic lattices}},\ }\href@noop {} {\bibfield  {journal}
  {\bibinfo  {journal} {Nat. Phys.}\ }\textbf {\bibinfo {volume} {10}},\
  \bibinfo {pages} {39} (\bibinfo {year} {2014})}\BibitemShut {NoStop}%
\bibitem [{\citenamefont {Lubensky}\ \emph {et~al.}(2015)\citenamefont
  {Lubensky}, \citenamefont {Kane}, \citenamefont {Mao}, \citenamefont
  {Souslov},\ and\ \citenamefont {Sun}}]{Lubensky-Review-2015}%
  \BibitemOpen
  \bibfield  {author} {\bibinfo {author} {\bibfnamefont {T.}~\bibnamefont
  {Lubensky}}, \bibinfo {author} {\bibfnamefont {C.}~\bibnamefont {Kane}},
  \bibinfo {author} {\bibfnamefont {X.}~\bibnamefont {Mao}}, \bibinfo {author}
  {\bibfnamefont {A.}~\bibnamefont {Souslov}},\ and\ \bibinfo {author}
  {\bibfnamefont {K.}~\bibnamefont {Sun}},\ }\bibfield  {title} {\bibinfo
  {title} {{Phonons and elasticity in critically coordinated lattices}},\
  }\href@noop {} {\bibfield  {journal} {\bibinfo  {journal} {Rep. Prog. Phys.}\
  }\textbf {\bibinfo {volume} {78}},\ \bibinfo {pages} {073901} (\bibinfo
  {year} {2015})}\BibitemShut {NoStop}%
\bibitem [{\citenamefont {Pellegrino}\ and\ \citenamefont
  {Calladine}(1986)}]{Pellegrino-IJSS-22}%
  \BibitemOpen
  \bibfield  {author} {\bibinfo {author} {\bibfnamefont {S.}~\bibnamefont
  {Pellegrino}}\ and\ \bibinfo {author} {\bibfnamefont {C.~R.}\ \bibnamefont
  {Calladine}},\ }\bibfield  {title} {\bibinfo {title} {{Matrix analysis of
  statically and kinematically indeterminate frameworks}},\ }\href@noop {}
  {\bibfield  {journal} {\bibinfo  {journal} {Int. J. Solid Struct.}\ }\textbf
  {\bibinfo {volume} {22}},\ \bibinfo {pages} {409} (\bibinfo {year}
  {1986})}\BibitemShut {NoStop}%
\bibitem [{\citenamefont {Pellegrino}(1990)}]{Pellegrino-IJSS-1990}%
  \BibitemOpen
  \bibfield  {author} {\bibinfo {author} {\bibfnamefont {S.}~\bibnamefont
  {Pellegrino}},\ }\bibfield  {title} {\bibinfo {title} {{Analysis of
  prestressed mechanisms}},\ }\href@noop {} {\bibfield  {journal} {\bibinfo
  {journal} {Int. J. Solid Struct.}\ }\textbf {\bibinfo {volume} {26}},\
  \bibinfo {pages} {1329} (\bibinfo {year} {1990})}\BibitemShut {NoStop}%
\bibitem [{\citenamefont {Guest}(2006)}]{Guest-IJSS-2005}%
  \BibitemOpen
  \bibfield  {author} {\bibinfo {author} {\bibfnamefont {S.}~\bibnamefont
  {Guest}},\ }\bibfield  {title} {\bibinfo {title} {{The stiffness of
  prestressed frameworks: A unifying approach}},\ }\href@noop {} {\bibfield
  {journal} {\bibinfo  {journal} {Int. J. Solid Struct.}\ }\textbf {\bibinfo
  {volume} {43}},\ \bibinfo {pages} {842} (\bibinfo {year} {2006})}\BibitemShut
  {NoStop}%
\bibitem [{\citenamefont {ge~Chen}\ \emph {et~al.}(2014)\citenamefont
  {ge~Chen}, \citenamefont {Upadhyaya},\ and\ \citenamefont
  {Vitelli}}]{Chen-PNAS-2014}%
  \BibitemOpen
  \bibfield  {author} {\bibinfo {author} {\bibfnamefont {B.~G.}\ \bibnamefont
  {ge~Chen}}, \bibinfo {author} {\bibfnamefont {N.}~\bibnamefont {Upadhyaya}},\
  and\ \bibinfo {author} {\bibfnamefont {V.}~\bibnamefont {Vitelli}},\
  }\bibfield  {title} {\bibinfo {title} {{Nonlinear conduction via solitons in
  a topological mechanical insulator}},\ }\href@noop {} {\bibfield  {journal}
  {\bibinfo  {journal} {Proc. Nat. Acad. Sci. USA}\ }\textbf {\bibinfo {volume}
  {111}},\ \bibinfo {pages} {13004} (\bibinfo {year} {2014})}\BibitemShut
  {NoStop}%
\bibitem [{com()}]{comment-unpublished}%
  \BibitemOpen
  \href@noop {} {}\bibinfo {note} {{The chainmail-based meta-structures, such
  as those stuided in Ref.~\cite{Klotz-SoftMat-2024}, feature intriguing
  geometry and mechanics, which we will report
  elsewhere~\cite{unpublished}.}}\BibitemShut {Stop}%
\bibitem [{\citenamefont {Deng}\ \emph {et~al.}(2021)\citenamefont {Deng},
  \citenamefont {Raney}, \citenamefont {Bertoldi},\ and\ \citenamefont
  {Tournat}}]{Deng-JAP-2021}%
  \BibitemOpen
  \bibfield  {author} {\bibinfo {author} {\bibfnamefont {B.}~\bibnamefont
  {Deng}}, \bibinfo {author} {\bibfnamefont {J.~R.}\ \bibnamefont {Raney}},
  \bibinfo {author} {\bibfnamefont {K.}~\bibnamefont {Bertoldi}},\ and\
  \bibinfo {author} {\bibfnamefont {V.}~\bibnamefont {Tournat}},\ }\bibfield
  {title} {\bibinfo {title} {{Nonlinear waves in flexible mechanical
  metamaterials}},\ }\href@noop {} {\bibfield  {journal} {\bibinfo  {journal}
  {J. Appl. Phys.}\ }\textbf {\bibinfo {volume} {130}},\ \bibinfo {pages}
  {040901} (\bibinfo {year} {2021})}\BibitemShut {NoStop}%
\bibitem [{\citenamefont {Yasuda}\ \emph {et~al.}(2019)\citenamefont {Yasuda},
  \citenamefont {Miyazawa}, \citenamefont {Charalampidis}, \citenamefont
  {Chong}, \citenamefont {Kevrekidis},\ and\ \citenamefont
  {Yang}}]{Yasuda-SciAdv-2019}%
  \BibitemOpen
  \bibfield  {author} {\bibinfo {author} {\bibfnamefont {H.}~\bibnamefont
  {Yasuda}}, \bibinfo {author} {\bibfnamefont {Y.}~\bibnamefont {Miyazawa}},
  \bibinfo {author} {\bibfnamefont {E.~G.}\ \bibnamefont {Charalampidis}},
  \bibinfo {author} {\bibfnamefont {C.}~\bibnamefont {Chong}}, \bibinfo
  {author} {\bibfnamefont {P.~G.}\ \bibnamefont {Kevrekidis}},\ and\ \bibinfo
  {author} {\bibfnamefont {J.}~\bibnamefont {Yang}},\ }\bibfield  {title}
  {\bibinfo {title} {{Origami-based impact mitigation via rarefaction solitary
  wave creation}},\ }\href@noop {} {\bibfield  {journal} {\bibinfo  {journal}
  {Sci. Adv.}\ }\textbf {\bibinfo {volume} {5}},\ \bibinfo {pages} {eaau2835}
  (\bibinfo {year} {2019})}\BibitemShut {NoStop}%
\bibitem [{\citenamefont {\textcolor{red}{R. Imada and T.
  Tachi}}(2025)}]{Imada-PRRes-2025}%
  \BibitemOpen
  \bibfield  {author} {\bibinfo {author} {\bibnamefont {\textcolor{red}{R.
  Imada and T. Tachi}}},\ }\bibfield  {title} {\bibinfo {title}
  {\textcolor{red}{Maxwell origami tube}},\ }\href@noop {} {\bibfield
  {journal} {\bibinfo  {journal} {\textcolor{red}{Phys. Rev. Research}}\
  }\textbf {\bibinfo {volume} {\textcolor{red}{7}}},\ \bibinfo {pages}
  {\textcolor{red}{013032}} (\bibinfo {year}
  {\textcolor{red}{2025}})}\BibitemShut {NoStop}%
\bibitem [{\citenamefont {Deng}\ \emph {et~al.}(2017)\citenamefont {Deng},
  \citenamefont {Raney}, \citenamefont {Tournat},\ and\ \citenamefont
  {Bertoldi}}]{Deng-PRL-2017}%
  \BibitemOpen
  \bibfield  {author} {\bibinfo {author} {\bibfnamefont {B.}~\bibnamefont
  {Deng}}, \bibinfo {author} {\bibfnamefont {J.~R.}\ \bibnamefont {Raney}},
  \bibinfo {author} {\bibfnamefont {V.}~\bibnamefont {Tournat}},\ and\ \bibinfo
  {author} {\bibfnamefont {K.}~\bibnamefont {Bertoldi}},\ }\bibfield  {title}
  {\bibinfo {title} {{Elastic Vector Solitons in Soft Architeched Materials}},\
  }\href@noop {} {\bibfield  {journal} {\bibinfo  {journal} {Phys. Rev. Lett.}\
  }\textbf {\bibinfo {volume} {118}},\ \bibinfo {pages} {204102} (\bibinfo
  {year} {2017})}\BibitemShut {NoStop}%
\bibitem [{\citenamefont {Nadkarni}\ \emph {et~al.}(2016)\citenamefont
  {Nadkarni}, \citenamefont {Arrieta}, \citenamefont {Chong}, \citenamefont
  {Kochmann},\ and\ \citenamefont {Daraio}}]{Nadkarni-PRL-2016}%
  \BibitemOpen
  \bibfield  {author} {\bibinfo {author} {\bibfnamefont {N.}~\bibnamefont
  {Nadkarni}}, \bibinfo {author} {\bibfnamefont {A.~F.}\ \bibnamefont
  {Arrieta}}, \bibinfo {author} {\bibfnamefont {C.}~\bibnamefont {Chong}},
  \bibinfo {author} {\bibfnamefont {D.~M.}\ \bibnamefont {Kochmann}},\ and\
  \bibinfo {author} {\bibfnamefont {C.}~\bibnamefont {Daraio}},\ }\bibfield
  {title} {\bibinfo {title} {{Unidirectional Transition Waves in Bistable
  Lattices}},\ }\href@noop {} {\bibfield  {journal} {\bibinfo  {journal} {Phys.
  Rev. Lett.}\ }\textbf {\bibinfo {volume} {116}},\ \bibinfo {pages} {244501}
  (\bibinfo {year} {2016})}\BibitemShut {NoStop}%
\bibitem [{\citenamefont {Boucher}\ \emph {et~al.}(2017)\citenamefont
  {Boucher}, \citenamefont {Clanet}, \citenamefont {Qu\'{e}r\'{e}},\ and\
  \citenamefont {Chevy}}]{Boucher-PRL-2017}%
  \BibitemOpen
  \bibfield  {author} {\bibinfo {author} {\bibfnamefont {J.-P.}\ \bibnamefont
  {Boucher}}, \bibinfo {author} {\bibfnamefont {C.}~\bibnamefont {Clanet}},
  \bibinfo {author} {\bibfnamefont {D.}~\bibnamefont {Qu\'{e}r\'{e}}},\ and\
  \bibinfo {author} {\bibfnamefont {F.}~\bibnamefont {Chevy}},\ }\bibfield
  {title} {\bibinfo {title} {{Popsicle-Stick Cobra Wave}},\ }\href@noop {}
  {\bibfield  {journal} {\bibinfo  {journal} {Phys. Rev. Lett.}\ }\textbf
  {\bibinfo {volume} {119}},\ \bibinfo {pages} {084301} (\bibinfo {year}
  {2017})}\BibitemShut {NoStop}%
\bibitem [{\citenamefont {Veenstra}\ \emph {et~al.}(2024)\citenamefont
  {Veenstra}, \citenamefont {Gamayun}, \citenamefont {Guo}, \citenamefont
  {Sarvi}, \citenamefont {Meinersen},\ and\ \citenamefont
  {Coulais}}]{Veenstra-Nature-2024}%
  \BibitemOpen
  \bibfield  {author} {\bibinfo {author} {\bibfnamefont {J.}~\bibnamefont
  {Veenstra}}, \bibinfo {author} {\bibfnamefont {O.}~\bibnamefont {Gamayun}},
  \bibinfo {author} {\bibfnamefont {X.}~\bibnamefont {Guo}}, \bibinfo {author}
  {\bibfnamefont {A.}~\bibnamefont {Sarvi}}, \bibinfo {author} {\bibfnamefont
  {C.~V.}\ \bibnamefont {Meinersen}},\ and\ \bibinfo {author} {\bibfnamefont
  {C.}~\bibnamefont {Coulais}},\ }\bibfield  {title} {\bibinfo {title}
  {{Non-reciprocal topological solitons in active metamaterials}},\ }\href@noop
  {} {\bibfield  {journal} {\bibinfo  {journal} {Nature}\ }\textbf {\bibinfo
  {volume} {627}},\ \bibinfo {pages} {528} (\bibinfo {year}
  {2024})}\BibitemShut {NoStop}%
\bibitem [{\citenamefont {Scott}(1969)}]{Scott-AmJPhys-1969}%
  \BibitemOpen
  \bibfield  {author} {\bibinfo {author} {\bibfnamefont {A.~C.}\ \bibnamefont
  {Scott}},\ }\bibfield  {title} {\bibinfo {title} {{A Nonlinear Klein-Gordon
  Equation}},\ }\href@noop {} {\bibfield  {journal} {\bibinfo  {journal} {Am.
  J. Phys.}\ }\textbf {\bibinfo {volume} {37}},\ \bibinfo {pages} {52}
  (\bibinfo {year} {1969})}\BibitemShut {NoStop}%
\bibitem [{\citenamefont {Toda}(1989)}]{Toda-book}%
  \BibitemOpen
  \bibfield  {author} {\bibinfo {author} {\bibfnamefont {M.}~\bibnamefont
  {Toda}},\ }\href@noop {} {\emph {\bibinfo {title} {{Theory of Nonlinear
  Lattices}}}}\ (\bibinfo  {publisher} {Springer-Verlag},\ \bibinfo {address}
  {Heidelberg},\ \bibinfo {year} {1989})\BibitemShut {NoStop}%
\bibitem [{\citenamefont {Berman}\ and\ \citenamefont
  {Izrailev}(2005)}]{Berman-Chaos-2005}%
  \BibitemOpen
  \bibfield  {author} {\bibinfo {author} {\bibfnamefont {G.~P.}\ \bibnamefont
  {Berman}}\ and\ \bibinfo {author} {\bibfnamefont {F.~M.}\ \bibnamefont
  {Izrailev}},\ }\bibfield  {title} {\bibinfo {title} {{The Fermi-Pasta-Ulam
  problem: Fifty-years of progress}},\ }\href@noop {} {\bibfield  {journal}
  {\bibinfo  {journal} {Chaos}\ }\textbf {\bibinfo {volume} {15}},\ \bibinfo
  {pages} {015104} (\bibinfo {year} {2005})}\BibitemShut {NoStop}%
\bibitem [{\citenamefont {Kawasaki}\ and\ \citenamefont
  {Ohta}(1982)}]{Kawasaki-PhysicaA-1982}%
  \BibitemOpen
  \bibfield  {author} {\bibinfo {author} {\bibfnamefont {K.}~\bibnamefont
  {Kawasaki}}\ and\ \bibinfo {author} {\bibfnamefont {T.}~\bibnamefont
  {Ohta}},\ }\bibfield  {title} {\bibinfo {title} {{Kink Dynamics in
  One-dimensional Nonlinear Systems}},\ }\href@noop {} {\bibfield  {journal}
  {\bibinfo  {journal} {Physica}\ }\textbf {\bibinfo {volume} {116A}},\
  \bibinfo {pages} {573} (\bibinfo {year} {1982})}\BibitemShut {NoStop}%
\bibitem [{\citenamefont {Sekimoto}(2024)}]{Sekimoto-EPL-2024}%
  \BibitemOpen
  \bibfield  {author} {\bibinfo {author} {\bibfnamefont {K.}~\bibnamefont
  {Sekimoto}},\ }\bibfield  {title} {\bibinfo {title} {{Allosteric propagation
  of curvature along filament}},\ }\href@noop {} {\bibfield  {journal}
  {\bibinfo  {journal} {EPL}\ }\textbf {\bibinfo {volume} {147}},\ \bibinfo
  {pages} {60001} (\bibinfo {year} {2024})}\BibitemShut {NoStop}%
\bibitem [{\citenamefont {Goldstein}\ \emph {et~al.}(2000)\citenamefont
  {Goldstein}, \citenamefont {Goriely}, \citenamefont {Huber},\ and\
  \citenamefont {Wolgemuth}}]{Goldstein-PRL-2000}%
  \BibitemOpen
  \bibfield  {author} {\bibinfo {author} {\bibfnamefont {R.~E.}\ \bibnamefont
  {Goldstein}}, \bibinfo {author} {\bibfnamefont {A.}~\bibnamefont {Goriely}},
  \bibinfo {author} {\bibfnamefont {G.}~\bibnamefont {Huber}},\ and\ \bibinfo
  {author} {\bibfnamefont {C.~W.}\ \bibnamefont {Wolgemuth}},\ }\bibfield
  {title} {\bibinfo {title} {{Bistable Helices}},\ }\href@noop {} {\bibfield
  {journal} {\bibinfo  {journal} {Phys. Rev. Lett.}\ }\textbf {\bibinfo
  {volume} {84}},\ \bibinfo {pages} {1631} (\bibinfo {year}
  {2000})}\BibitemShut {NoStop}%
\bibitem [{\citenamefont {Shaevitz}\ \emph {et~al.}(2005)\citenamefont
  {Shaevitz}, \citenamefont {Lee},\ and\ \citenamefont
  {Fletcher}}]{Shaevitz-Cell-2005}%
  \BibitemOpen
  \bibfield  {author} {\bibinfo {author} {\bibfnamefont {J.~W.}\ \bibnamefont
  {Shaevitz}}, \bibinfo {author} {\bibfnamefont {J.~Y.}\ \bibnamefont {Lee}},\
  and\ \bibinfo {author} {\bibfnamefont {D.~A.}\ \bibnamefont {Fletcher}},\
  }\bibfield  {title} {\bibinfo {title} {{{\it Spiroplasma} Swim by a
  Processive Change in Body Helicity}},\ }\href@noop {} {\bibfield  {journal}
  {\bibinfo  {journal} {Cell}\ }\textbf {\bibinfo {volume} {122}},\ \bibinfo
  {pages} {941} (\bibinfo {year} {2005})}\BibitemShut {NoStop}%
\bibitem [{\citenamefont {Nakane}\ \emph {et~al.}(2020)\citenamefont {Nakane},
  \citenamefont {Ito},\ and\ \citenamefont
  {Nishizaka}}]{Nakane-JBacteriol-2020}%
  \BibitemOpen
  \bibfield  {author} {\bibinfo {author} {\bibfnamefont {D.}~\bibnamefont
  {Nakane}}, \bibinfo {author} {\bibfnamefont {T.}~\bibnamefont {Ito}},\ and\
  \bibinfo {author} {\bibfnamefont {T.}~\bibnamefont {Nishizaka}},\ }\bibfield
  {title} {\bibinfo {title} {{Coexistence of Two Chiral Helices Produces Kink
  Translation in {\it Spiroplasma} Swimming}},\ }\href@noop {} {\bibfield
  {journal} {\bibinfo  {journal} {J. Bacteriol.}\ }\textbf {\bibinfo {volume}
  {202}},\ \bibinfo {pages} {e00735} (\bibinfo {year} {2020})}\BibitemShut
  {NoStop}%
\bibitem [{\citenamefont {Sasajima}\ and\ \citenamefont
  {Miyata}(2021)}]{Sasajima-FrontBiol-2021}%
  \BibitemOpen
  \bibfield  {author} {\bibinfo {author} {\bibfnamefont {Y.}~\bibnamefont
  {Sasajima}}\ and\ \bibinfo {author} {\bibfnamefont {M.}~\bibnamefont
  {Miyata}},\ }\bibfield  {title} {\bibinfo {title} {{Prospects for the
  Mechanism of {\it Spiroplasma} Swimming}},\ }\href@noop {} {\bibfield
  {journal} {\bibinfo  {journal} {Front. Microbiol.}\ }\textbf {\bibinfo
  {volume} {12}},\ \bibinfo {pages} {706426} (\bibinfo {year}
  {2021})}\BibitemShut {NoStop}%
\bibitem [{\citenamefont {Klotz}\ \emph {et~al.}(2024)\citenamefont {Klotz},
  \citenamefont {Anderson},\ and\ \citenamefont
  {Dimitriyev}}]{Klotz-SoftMat-2024}%
  \BibitemOpen
  \bibfield  {author} {\bibinfo {author} {\bibfnamefont {A.~R.}\ \bibnamefont
  {Klotz}}, \bibinfo {author} {\bibfnamefont {C.~J.}\ \bibnamefont
  {Anderson}},\ and\ \bibinfo {author} {\bibfnamefont {M.~S.}\ \bibnamefont
  {Dimitriyev}},\ }\bibfield  {title} {\bibinfo {title} {{Chirality effects in
  molecular chainmail}},\ }\href@noop {} {\bibfield  {journal} {\bibinfo
  {journal} {Soft Matter}\ }\textbf {\bibinfo {volume} {20}},\ \bibinfo {pages}
  {7044} (\bibinfo {year} {2024})}\BibitemShut {NoStop}%
\bibitem [{\citenamefont {Ueno}\ \emph {et~al.}(2026)\citenamefont {Ueno},
  \citenamefont {Shimamoto}, \citenamefont {Yoneda},\ and\ \citenamefont
  {Wada}}]{unpublished}%
  \BibitemOpen
  \bibfield  {author} {\bibinfo {author} {\bibfnamefont {S.}~\bibnamefont
  {Ueno}}, \bibinfo {author} {\bibfnamefont {D.}~\bibnamefont {Shimamoto}},
  \bibinfo {author} {\bibfnamefont {T.}~\bibnamefont {Yoneda}},\ and\ \bibinfo
  {author} {\bibfnamefont {H.}~\bibnamefont {Wada}},\ }\bibfield  {title}
  {\bibinfo {title} {Topological waves in chainmail metasheets}} (\bibinfo
  {year} {2026}),\ \bibinfo {note} {to be published}\BibitemShut {NoStop}%
\bibitem [{\citenamefont {Kamada}\ and\ \citenamefont
  {Yasuda}(1998)}]{Karakuri-Book}%
  \BibitemOpen
  \bibfield  {author} {\bibinfo {author} {\bibfnamefont {M.}~\bibnamefont
  {Kamada}}\ and\ \bibinfo {author} {\bibfnamefont {M.}~\bibnamefont
  {Yasuda}},\ }\href@noop {} {\emph {\bibinfo {title} {{Let's Make Antiqued
  Mechanical Toys (in Japanese)}}}}\ (\bibinfo  {publisher} {Kawade Shobo
  Shinsha},\ \bibinfo {year} {1998})\BibitemShut {NoStop}%
\end{thebibliography}
\end{document}